%% file: arxiv.tex
\documentclass[numbers,sort&compress]{article}
\usepackage{graphicx}
\usepackage{rotating}
\usepackage[normalem]{ulem}
\usepackage{arxiv}
\usepackage{amssymb,amsmath,amsthm}
\usepackage{natbib}
\usepackage{xcolor}
\usepackage{hyperref}
\usepackage[toc,page]{appendix}
\graphicspath{{Artworks/}}
\graphicspath{{imgs/}}
\usepackage{algorithm2e}
\theoremstyle{definition}
\usepackage{url}            %
\usepackage{booktabs}       %
\usepackage{amsfonts}       %
\usepackage{nicefrac}       %
\usepackage{microtype}      %
\usepackage{cleveref}       %
\usepackage{graphicx}
\usepackage{doi}
\input{authors_arxiv.tex}
\date{\today}

\title{Bayesian Hierarchical Models for Quantitative Estimates for Performance metrics applied to Saddle Search Algorithms}
\hypersetup{
 pdfauthor={Rohit Goswami (HaoZeke)},
 pdftitle={Bayesian Hierarchical Models for Quantitative Estimates for Performance metrics applied to Saddle Search Algorithms},
 pdfkeywords={},
 pdfsubject={},
 pdfcreator={Emacs 30.1 (Org mode 9.7.31)}, 
 pdflang={English}}
\begin{document}

\maketitle
\begin{abstract} %
Rigorous performance evaluation is essential for developing robust algorithms for high-throughput computational chemistry. Traditional benchmarking, however, often struggles to account for system-specific variability, making it difficult to form actionable conclusions. We present a Bayesian hierarchical modeling framework that rigorously quantifies performance metrics and their uncertainty, enabling a nuanced comparison of algorithmic strategies. We apply this framework to analyze the Dimer method, comparing Conjugate Gradient (CG) and L-BFGS rotation optimizers, with and without the removal of external rotations, across a benchmark of 500 molecular systems. Our analysis confirms that CG offers higher overall robustness than L-BFGS in this context. While the theoretically-motivated removal of external rotations led to higher computational cost (>40\% more energy and force calls) for most systems in this set, our models also reveal a subtle interplay, hinting that this feature may improve the reliability of the L-BFGS optimizer. Rather than identifying a single superior method, our findings support the design of adaptive ``chain of methods'' workflows. This work showcases how a robust statistical paradigm can move beyond simple performance rankings to inform the intelligent, context-dependent application of computational chemistry methods.
\end{abstract}
\section{Introduction}
\label{sec:introduction}
Locating transition states (TS), first-order saddle points on potential energy surfaces (PES), is fundamental to understanding reaction mechanisms and kinetics in chemistry \cite{eyringActivatedComplexChemical1935,pechukasRecentDevelopmentsTransition1982}. Numerous algorithms target this challenge, with minimum mode following (MMF) methods forming a prominent class \cite{cerjanFindingTransitionStates1981}. These algorithms, initiated from a single point on the PES, iteratively ascend the minimum curvature mode while relaxing orthogonal directions. Notable MMF families include Lanczos-based Activation Relaxation Technique Nouveau (ARTn) methods \cite{mousseauTravelingPotentialEnergy1998,cancesImprovementsActivationrelaxationTechnique2009,mousseauActivationRelaxationTechniqueART2012,poberznikPARTnPluginImplementation2024,gundeExploringPotentialEnergy2024} and the two-point Dimer method \cite{henkelmanDimerMethodFinding1999}, which uses a pair of images to estimate the minimum mode direction. This is in turn an independent implementation of the hybrid eigenvector-following algorithm \cite{munroDefectMigrationCrystalline1999}. Other variations employ internal coordinates \cite{hermesSellaOpenSourceAutomationFriendly2022} or iteratively build PES approximations \cite{koistinenMinimumModeSaddle2020,goswamiEfficientImplementationGaussian2025a}.

Evaluating and comparing such algorithms typically relies on benchmarks, from established sets like Baker's \cite{bakerLocationTransitionStates1996} to more recent collections \cite{chillBenchmarksCharacterizationMinima2014}. However, studies introducing methodological advancements often demonstrate performance on limited sets of small systems, frequently using computationally inexpensive potentials \cite{koistinenMinimumModeSaddle2020,melanderRemovingExternalDegrees2015,zengUnificationAlgorithmsMinimum2014,kastnerSuperlinearlyConvergingDimer2008,olsenComparisonMethodsFinding2004,lengEfficientSoftestMode2013}. This traditional benchmarking paradigm appears increasingly misaligned with current computational practice, where peta- and exa-scale resources combined with workflow engines \cite{huberAutomatedReproducibleWorkflows2022,molderSustainableDataAnalysis2021} enable large-scale, systematic explorations across diverse chemical spaces. Simple performance comparisons based on small benchmarks or average metrics often neglect significant system-to-system variability and potential implementation biases \cite{kriegelBlackArtRuntime2017,flemingHowNotLie1986}, providing an inadequate basis for selecting optimal methods for demanding high-throughput applications.

To bridge this gap, we advocate for and demonstrate a rigorous statistical framework using Bayesian generalized linear mixed-effects models (GLMMs) for analyzing algorithm performance data from large benchmark sets. This approach, implemented via the \texttt{brms} R package \cite{burknerBrmsPackageBayesian2017} interfacing with Stan \cite{carpenterStanProbabilisticProgramming2017}, robustly handles the statistical modeling of various performance metrics (e.g., computation time, PES call counts, binary success) and explicitly models the hierarchical structure of benchmark data (e.g., multiple runs within chemical systems) using random effects. Crucially, the Bayesian paradigm provides comprehensive uncertainty quantification through full posterior distributions and credible intervals, facilitating nuanced and reliable comparisons beyond simple point estimates or visual inspection.

In this work, we apply this Bayesian GLMM framework to analyze the performance trade-offs associated with two common variations within the Dimer method: the choice between Conjugate Gradient (CG) and Limited-memory BFGS (L-BFGS) optimizers for the rotational step \cite{kastnerSuperlinearlyConvergingDimer2008}, and the effect of enabling or disabling the removal of external rotation and translation degrees of freedom \cite{melanderRemovingExternalDegrees2015}. We utilize the EON software package \cite{chillEONSoftwareLong2014} interfaced with NWChem \cite{apraNWChemPresentFuture2020} at the HF/3-21G level of theory, analyzing performance across a diverse benchmark set of 500 initial configurations near saddle points introduced by Hermes et al. \cite{hermesSellaOpenSourceAutomationFriendly2022}.

The remainder of this paper is structured as follows: Section \ref{sec:methods} details the Dimer method variants and the statistical models employed. Section \ref{sec:appl} presents the results of applying these models to the benchmark dataset. Section \ref{sec:discussion} discusses the implications of the findings and the advantages of the statistical approach, and Section \ref{sec:concl} provides concluding remarks.
\section{Methods}
\label{sec:methods}
\subsection{Dimer method}
\label{sec:org3e0d45c}
The Dimer method \cite{henkelmanDimerMethodFinding1999} is a single-ended saddle search method that algorithmically finds first-order saddle point geometries by iteratively refining the position \(R\) (a 3N-dimensional vector representing the system coordinates) and orientation \(\hat{N}\) (a unit vector in \(3N\) dimensions) of a ``dimer''. An alternative formulation of the same was demonstrated earlier via the hybrid-eigenvector following using the variational Rayleigh-Ritz to find the eigenvalue and eigenvector pair \cite{munroDefectMigrationCrystalline1999}. The dimer consists of two system images, \(R_1\) and \(R_2\), separated by a small fixed distance \(\Delta R\) along the dimer axis \(\hat{N}\):
\begin{align}
    R_1 &= R - \frac{\Delta R}{2} \hat{N} \\
    R_2 &= R + \frac{\Delta R}{2} \hat{N}
\end{align}
The core idea is to rotate the dimer orientation \(\hat{N}\) to align with the lowest curvature mode at the midpoint \(R\), and then translate \(R\) uphill along this mode (if curvature is negative) while minimizing energy in all other directions. This process typically involves two main steps per iteration: rotation and translation.

The goal of the rotation step is to find the orientation \(\hat{N}\) that minimizes the dimer's energy, keeping the midpoint \(R\) fixed. This corresponds to aligning \(\hat{N}\) with the direction of lowest local curvature. The rotational dynamics are driven by the component of the sum of forces at the endpoints perpendicular to the dimer axis, \(F_{rot} = (F_1 + F_2) - ((F_1 + F_2) \cdot \hat{N}) \hat{N}\), where \(F_1 = F(R_1)\) and \(F_2 = F(R_2)\). The rotation aims to drive \(F_{rot}\) to zero.

Calculating the exact energy curvature requires the Hessian matrix. To avoid this computational expense, the Dimer method typically employs approximations based only on energies and forces at the endpoints \(R_1\) and \(R_2\). Following the approach refined by Olsen et al. \cite{olsenComparisonMethodsFinding2004}, the curvature along the dimer axis \(C(\hat{N})\) and an effective torque or rotational force \(\tau(\hat{N})\) can be estimated efficiently and iteratively until a threshold for the rotational force is reached. The finite difference approximation for the curvature is:
\begin{equation}
    C(\hat{N}) \approx \frac{(F_2 - F_1) \cdot \hat{N}}{\Delta R}
    \label{eq:dimer_curvature}
\end{equation}

The rotational force \(\tau(\hat{N})\), which is perpendicular to \(\hat{N}\), can also be estimated from \(F_1\) and \(F_2\) \cite{heydenEfficientMethodsFinding2005}. An optimization algorithm, such as Conjugate Gradient (CG) or Limited-memory BFGS (L-BFGS) as investigated in this work, is then used to iteratively update the orientation \(\hat{N}\) (typically by small rotations) to minimize the energy with respect to rotation, effectively driving \(\tau(\hat{N})\) towards zero.

Once the dimer is aligned with the estimated lowest mode direction \(\hat{N}\), the midpoint \(R\) is moved towards the saddle point. The translation is guided by an effective force \(F_{trans}\). If the estimated curvature \(C(\hat{N})\) is negative (indicating an uphill direction towards the saddle), the force component parallel to \(\hat{N}\) is inverted. The effective force for the optimizer is:
\begin{equation}
    F_{trans}(R) = F(R) - 2 (F(R) \cdot \hat{N}) \hat{N} \quad \text{if } C(\hat{N}) < 0
    \label{eq:dimer_trans_force}
\end{equation}

If \(C(\hat{N}) \ge 0\), the unmodified force \(F(R)\) might be used, or other strategies employed to ensure movement towards a saddle point. A standard optimization algorithm (often L-BFGS in many implementations) then takes a step using this effective force to update the midpoint:
\begin{equation}
    R_{k+1} = \text{OptimizerStep}(R_k, F_{trans}(R_k))
    \label{eq:dimer_trans_step}
\end{equation}

The rotation and translation steps are repeated iteratively until convergence criteria, typically based on the magnitude of the total force \(F(R)\), are met.
\subsection{Statistical models}
\label{sec:statmodels}
Quantitative assessment of saddle point search algorithm performance requires
rigorous statistical methods, particularly when comparing multiple algorithms
across diverse chemical systems. While foundational studies have provided
crucial insights
\cite{bakerLocationTransitionStates1996,hermesSellaOpenSourceAutomationFriendly2022,koistinenMinimumModeSaddle2020,kastnerSuperlinearlyConvergingDimer2008},
analyses often rely on qualitative comparisons or the eyeball norm and summary
statistics based on relatively small datasets. Such approaches may not fully
capture performance variability or allow for robust statistical inference
regarding the significance and magnitude of performance differences. When larger
system sets are used they are often hampered by point-estimates of error like
the root mean square deviation
\cite{fuForcesAreNot2022,starkBenchmarkingMachineLearning2024}. It is important to
note that much of the saddle search literature for molecular systems makes
several approximations valid for small angles or well behaved portions of the
energy landscape, which may not hold in practice.

The number of potential energy surface (PES) calls (computational cost)
represents count data, total computation time is typically positive and
right-skewed, and search success is a binary outcome. These data types often
violate the assumptions inherent in standard linear models (e.g., normality,
homoscedasticity) \cite{oharaNotLogtransformCount2010}. Additionally, comparing
algorithm performance typically involves applying multiple methods to the same
set of chemical systems, resulting in repeated measures or clustered data.
Observations from the same system are not independent, and failing to account
for this structure can lead to inaccurate standard errors and potentially flawed
conclusions. Variability arising from different software implementations or
environments can also present challenges \cite{kriegelBlackArtRuntime2017},
underscoring the need for reproducible analyses.

To analyze the saddle search results, we focus on interpreting three algorithmic
concerns, the computational cost, the total time taken, and the success ratio.
These are considered in the context of varying the optimizer used in the
rotation phase (CG versus LBFS) and the effect of using rotation removal (yes versus no) thus
forming 4 sets of measurements on the 500 systems. We employed Bayesian
generalized linear mixed-effects models
\cite{mcelreathStatisticalRethinkingBayesian2020}. This framework is well-suited
to the data structure and research questions. GLMMs explicitly model the
appropriate response distribution for each outcome (counts, continuous time,
binary success) using link functions, while incorporating random effects to
account for the repeated measures structure (i.e., baseline differences between
systems). The Bayesian approach provides full posterior distributions for
parameter estimates, facilitating comprehensive uncertainty quantification via
credible intervals and enabling principled model comparison
\cite{burknerBrmsPackageBayesian2017}. We use \texttt{brms}
\cite{burknerBrmsPackageBayesian2017} to construct a Stan model
\cite{carpenterStanProbabilisticProgramming2017} which is compiled into C++ code
for analysis with the no U turn sampler (NUTS)
\cite{homanNoUturnSamplerAdaptively2014}. These models are so efficient that they
are run on a local machine with CPU configuration of 13th Gen Intel i7-1365U
(12) @ 5.200GHz with 30 GB of available memory. Details of model fitting, prior
specification, and convergence diagnostics are provided in the Supplementary
Information.
\subsubsection{Effect structures considered}
\label{sec:org91c7e65}

All statistical models employed shared a common approach to constructing the
linear predictor, denoted by \(\eta_{ij}\), which corresponds to observation \(i\) on system
\(j\). The expected value of the outcome (\(\mu_{ij}\)) is related to this linear
predictor via an appropriate link function \(g(\cdot)\), i.e., \(g(\mu_{ij}) =
\eta_{ij}\), with the specific link function and response distribution detailed
below for each outcome type.

Each linear predictor \(\eta_{ij}\) included a random intercepts \(u_j\), to account for the repeated measures within system \(j\), i.e. \(u_j \sim \text{Normal}(0, \sigma^2_u)\).

The fixed-effects component was varied systematically to address different analytical
questions. The specific fixed-effects structures examined were:

\begin{equation}
\text{Fixed Effects}_{ij} =
    \begin{cases}
        \beta_0 + \beta_{1} \text{DR}_{i(j))} \quad \text{(RotOptimizer)} \\
        \beta_0 + \beta_{2} \text{RR}_{i(j))} \quad \text{(RotRemoval)} \\
        \beta_0 + \beta_{1} \text{DR}_{i(j))} + \beta_{2} \text{RR}_{i(j))}  \\
        \quad + \beta_{3} (\text{DR}_{i(j))} \times \text{RR}_{i(j))}) \quad \text{(Full)}
    \end{cases}
\label{eq:fixed_effects_structures}
\end{equation}

Here, \(\beta_k\) represents fixed-effects coefficients (intercept and slopes),
\(\text{DR}_{i(j))}\) is an indicator variable for the dimer rotation optimizer used
(CG or L-BFGS) for observation \(i\) within system \(j\), and \(\text{RR}_{i(j))}\) is an indicator variable
for rotation removal ('no' or 'yes') within system \(j\). The \(\text{DR}_{i(j))} \times \text{RR}_{i(j))}\) term
represents their interaction.

The first two effect settings focus on the main effect of a single factor
(RotOptimizer or RotRemoval, respectively). When fitted to the complete dataset,
these estimate the effect of the included factor while marginalizing over the
levels of the excluded factor. Alternatively, as detailed in the Results for
specific targeted comparisons, these structures might be applied to subsets of
the data (e.g., comparing optimizers only when rotatons are removed).

The third effect setting considers the main effects of the dimer rotation
optimizer and the rotation removal. The fourth effect setting represents the
full interaction model. It simultaneously estimates the main effects of both
factors and their interaction term. This model generally utilizes data from all
four conditions per system (where available) to provide the most comprehensive
assessment of how the factors jointly influence the outcome.

This approach allows for both focused investigation of individual factor
effects, potentially mirroring simpler experimental designs, and a detailed
analysis of their interplay using the full statistical power of the dataset
within the multilevel framework. The specific response distribution and link
function for each outcome metric are detailed in the following sections.
\subsubsection{Modeling algorithmic efficiency}
\label{sec:orgf0669e6}

The primary figure of merit for saddle search algorithms, particularly when
introducing new methodologies, is the number of calls made to the underlying
potential energy surface (PES). This metric, referred throughout this manuscript
as ``PES calls,'' directly correlates with computational cost, especially when
high-level electronic structure methods are employed.

PES calls represent count data \cite{oharaNotLogtransformCount2010}, which
typically violate the assumptions of standard linear models, which assume the response variable is normally distributed. Beyond this, the
difficulty of applying frequentist measures of statistical significance like the
p-value for applied situations is fraught with controversy
\cite{mcshaneAbandonStatisticalSignificance2019,kuffnerWhyArePValues2019}. As
demonstrated in the Supplementary Information, even linear mixed-effects models,
which account for per-molecule variability, can produce misleading results when
applied directly to count data without appropriate transformations or
distributional assumptions.

For comparing PES calls, we employed Bayesian generalized linear mixed-effects
models (GLMMs) with a negative binomial response distribution with a logarithmic
link function, which has support on the non negative integers. This distribution
is appropriate for overdispersed count data like PES calls
\cite{oharaNotLogtransformCount2010}. We thus use:

\begin{equation}
\begin{aligned}
\text{PESCalls}_{ij} &\sim \text{NegativeBinomial}(\mu_{ij}, \phi) \\
\log(\mu_{ij}) &= \text{Fixed Effects}_{ij} + u_j
\end{aligned}
\label{eq:brms_pes}
\end{equation}

where \(\text{PESCalls}_{ij}\) represents the observed PES calls for the \(i^{th}\)
observation on system \(j\); \(\mu_{ij}\) is the expected value; \(\phi\) is the shape
parameter of the negative binomial distribution; Fixed effects Eq.
\ref{eq:fixed_effects_structures} comprise of the overall intercept and effect of
the method; and \(u_j\) is a random intercept for chemical system \(j\), accounting for the paired nature of
the data and modeled as \(u_j \sim \text{Normal}(0, \sigma^2_u)\).
\subsubsection{Modeling total time}
\label{sec:orgfa6be45}

Total time analysis for saddle search algorithms is often perilous, since the
bulk of compute intesive calculation are offloaded to external codes like VASP,
ORCA etc. which may have strong scaling and depend heavily on the specific
functional form of the energy function along with the runtime environment.
Additionally, compared to the cost of making PES calls, the time spent within
the saddle search algorithm is almost always negligible. Nevertheless, since
total wall time elapsed is of practical importance, it provides a valid
algorithmic constraint for decision making.

For estimating the effect on total computation time, a similar GLMM structure
was employed. Given that time is a continuous, positive, and often right-skewed
variable, we used a Gamma response distribution with a logarithmic link
function:

\begin{equation}
\begin{aligned}
\text{TotalTime}_{ij} &\sim \text{Gamma}(\mu_{ij}, \alpha) \\
\log(\mu_{ij}) &= \text{Fixed Effects}_{ij} + u_j
\end{aligned}
\label{eq:brms_time}
\end{equation}

where \(\text{TotalTime}_{ij}\) is the computation time for observation \(i\) on
system \(j\), \(\mu_{ij}\) is the expected time, \(\alpha\) is the shape parameter of
the Gamma distribution (inversely related to the variance when the variance is presented in terms of the squared mean). The
linear predictor \(\log(\mu_{ij})\) incorporates the fixed effects specified in
Eq. \ref{eq:fixed_effects_structures} and the system-specific random intercept
\(u_j\), which is modeled as \(u_j \sim \text{Normal}(0, \sigma^2_u)\).
\subsubsection{Modeling success probabilities}
\label{sec:org359d565}
To analyze the factors influencing the likelihood of a successful saddle point
search convergence, we fitted a Bayesian generalized linear mixed model
appropriate for binary outcomes. This involved using a Bernoulli response
distribution and a logit link function:

\begin{equation}
\begin{aligned}
\text{Success}_{ij} &\sim \text{Bernoulli}(p_{ij}) \\
\text{logit}(p_{ij}) &= \log\left( \frac{p_{ij}}{1-p_{ij}} \right) = \text{Fixed Effects}_{ij} + u_j
\end{aligned}
\label{eq:brms_success}
\end{equation}

Here, \(\text{Success}_{ij}\) is a binary indicator for observation \(i\) on system
\(j\), \(p_{ij}\) is the probability of success, \(\text{logit}(p_{ij})\) is the
log-odds of success, and the linear predictor incorporates the fixed effects
specified in Eq. \eqref{eq:fixed_effects_structures} and the system-specific
random intercept \(u_j\), which is modeled as \(u_{j} \sim \text{Normal}(0, \sigma^2_u)\).

This model allows us to estimate how the different factors influence the odds of
a successful search, while accounting for baseline differences in success rates
across systems.
\section{Applications}
\label{sec:appl}
To demonstrate the utility of leveraging a Bayesian approach to performance
statistices, we analyse the performance of the dimer method variants on the
dataset of 500 initial configurations for small gas-phase organic molecules
(7-25 atoms) introduced by \citet{hermesSellaOpenSourceAutomationFriendly2022}.
Saddle point searches were conducted using the EON software package
\cite{chillEONSoftwareLong2014} \footnote{Using the Github variant:
\url{https://github.com/TheochemUI/eOn}}, interfaced with NWChem
\cite{apraNWChemPresentFuture2020} as the quantum mechanical engine. Energies and
forces were calculated at the Hartree-Fock (HF) level of theory using RHF
(restricted HF) for singlets and UHF (unrestricted HF) for doublets
\cite{jensenIntroductionComputationalChemistry2017} with the 3-21G basis set.
Saddle searches algorithms were considered converged when the maximum component
of the atomic forces fell below 0.01 eV/Å. This rather loose tolerance was chosen to highlight the effect of rotation removal, as the step sizes decrease on approaching the saddle and are less likely to lead to spurious rotation. Additionally, tighter tolerances tend to put more focus on the parameters of the underlying computational engine which is not the primary goal here. Comprehensive details regarding
computational parameters, including self-consistent field thresholds, specific
EON settings, and interface configurations, are provided in the Supplementary
Information.

All models involved pairwise comparisons of dimer control modalities, using data
filtered to include only systems where both methods in the pair successfully
converged, and with the same saddle, with an energy difference of less than
0.01eV. Full interaction models are also fit. Since both PES calls and total
time are proxies for computational effort, only the PES calls are presented
here, with the total time model in the Supplementary.
\subsection{Data overview}
\label{sec:data_overview}
We applied the four dimer method variants (combinations of CG/L-BFGS rotation optimizers and enabling/disabling rotation removal) to the benchmark set of 500 initial configurations. An initial exploratory analysis of the performance metrics provides context for the subsequent Bayesian modeling, focusing here on the number of potential energy surface (PES) calls for successful runs and overall convergence success.

\begin{figure}[htbp]
\centering
\includegraphics[width=.9\linewidth]{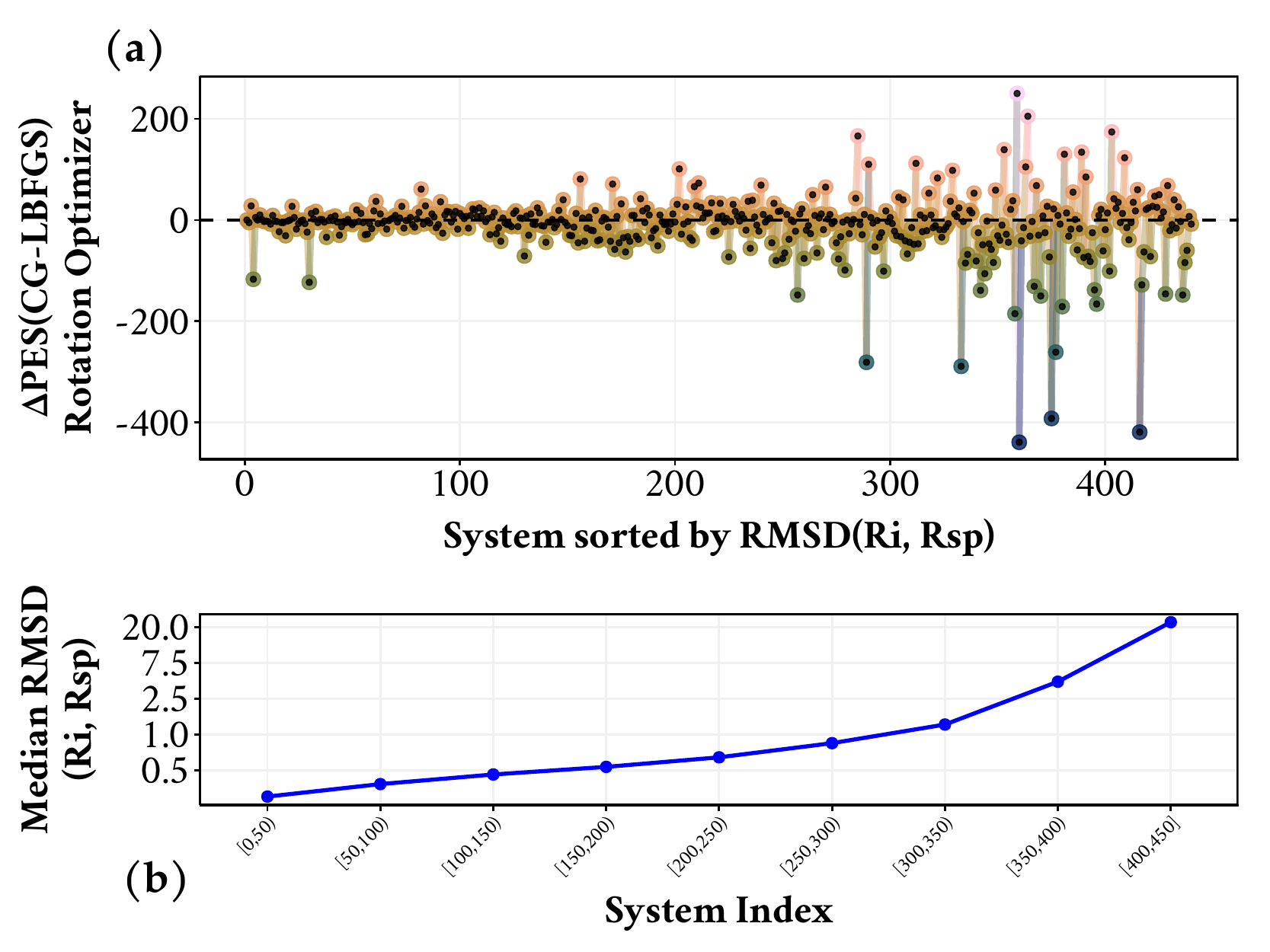}
\caption{\label{fig:rotoptcomp}(a) Difference in the number of potential energy surface (PES) calls between L-BFGS and conjugate gradient (CG) rotations for dimer calculations that converged to the same saddle, defined as having a final energy difference of less than 0.01 eV. The x-axis represents the system index, sorted by the median RMSD. The y-axis displays the difference in PES calls, for changing the optimizer for the rotation of the dimer when rotation removal is not used. Consequently, negative values indicate that the CG algorithm required fewer PES calls (i.e., CG performed better), while positive values indicate that L-BFGS required fewer PES calls (i.e., LBFGS performed better). The median improvement is 21 calls occuring in 55.23\% systems for CG-rotations compared to a median improvement of 16 calls in 43.18\% systems for LBFGS-rotations across 440 systems. Systems which are farther away from the saddle point show a larger improvement with CG-rotations. (b) For the systems shown in (a), their log(median RMSD) plotted against System Index, with Median RMSD displayed on the y-axis.}
\end{figure}

\begin{figure}[htbp]
\centering
\includegraphics[width=.9\linewidth]{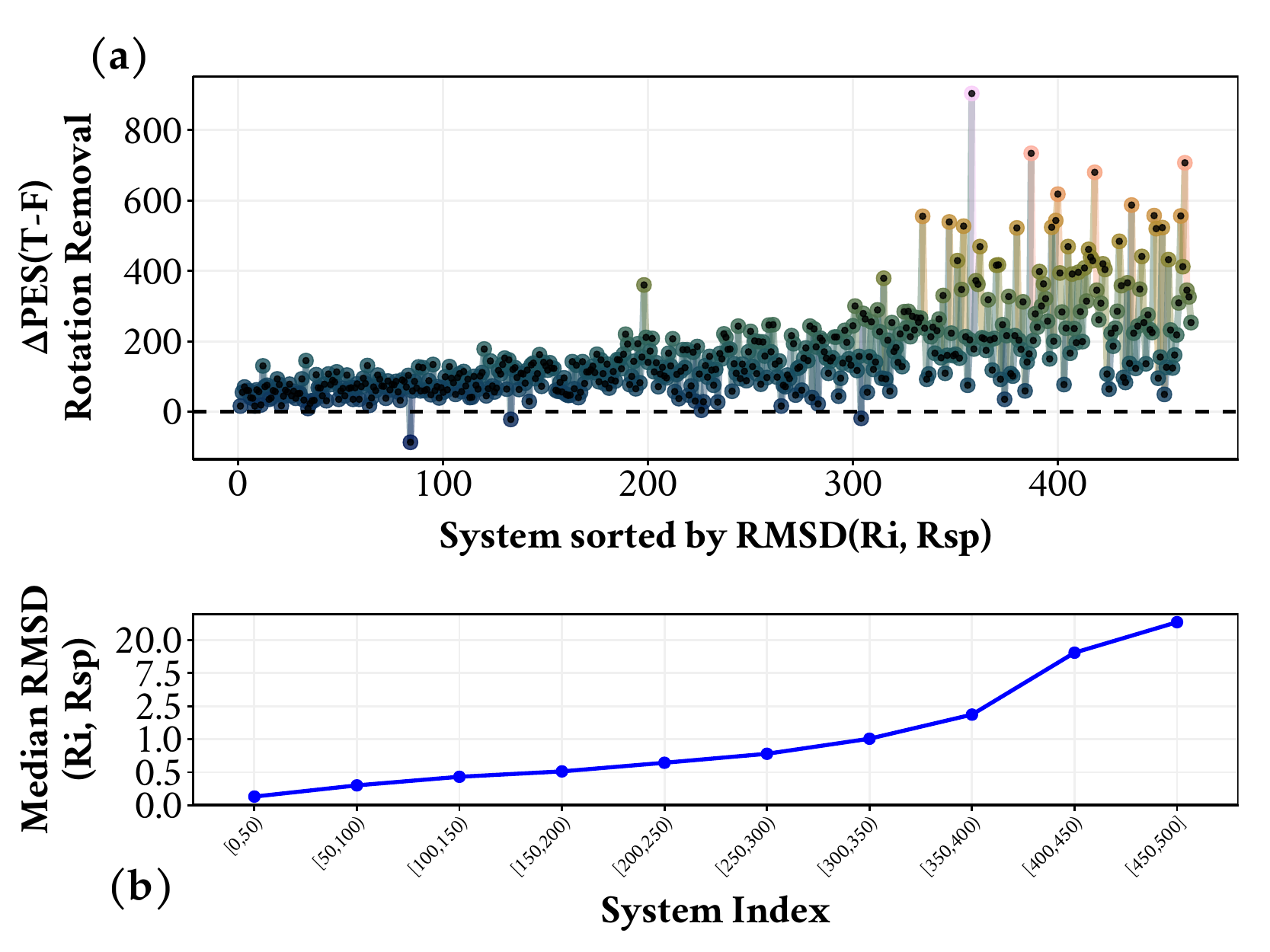}
\caption{\label{fig:rotremcomp}(a) Difference in the number of potential energy surface (PES) calls between using rotation removal (T) and not using rotation removal (F) for dimer calculations that converged to the same saddle, defined as having a final energy difference of less than 0.01 eV. The x-axis represents the system index, sorted by the median RMSD. The y-axis displays the difference in PES calls, using CG-rotations with and without rotation removal. Consequently, negative values indicate that the rotation removal required fewer PES calls (i.e., it performed better), while positive values indicate turning off the rotation removal took fewer PES calls (i.e., no rotation removal performed better). The median improvement is 126 calls occuring in 99.35\% systems for rotation removal not being used compared to a median improvement of 22 calls in 0.65\% systems when rotation removal is used across 465 systems. Systems which are farther away from the saddle point show a larger improvement without rotation removal. (b) For the systems shown in (a), their log(median RMSD) plotted against System Index, with Median RMSD displayed on the y-axis.}
\end{figure}

Figures \ref{fig:rotoptcomp} and \ref{fig:rotremcomp} illustrate the difference in PES calls required for convergence, comparing pairs of calculations that reached the same saddle point (within 0.01 eV). When rotation removal was disabled (Fig. \ref{fig:rotoptcomp}, 440 comparable systems), the CG-rotations dimer required fewer PES calls than L-BFGS-rotations in a majority of cases (55\%, median reduction 21 calls), although L-BFGS is rather comparable in performance (in 43\% cases, median reduction 16 calls). With CG-rotations (Fig. \ref{fig:rotremcomp}, 465 comparable systems), disabling rotation removal was markedly more efficient, requiring fewer PES calls than enabling it in over 99\% of systems (median reduction 126 calls). Both figures indicate that performance differences tend to be larger for systems starting with higher initial RMSD values relative to the saddle point.

\begin{figure}[htbp]
\centering
\includegraphics[width=.9\linewidth]{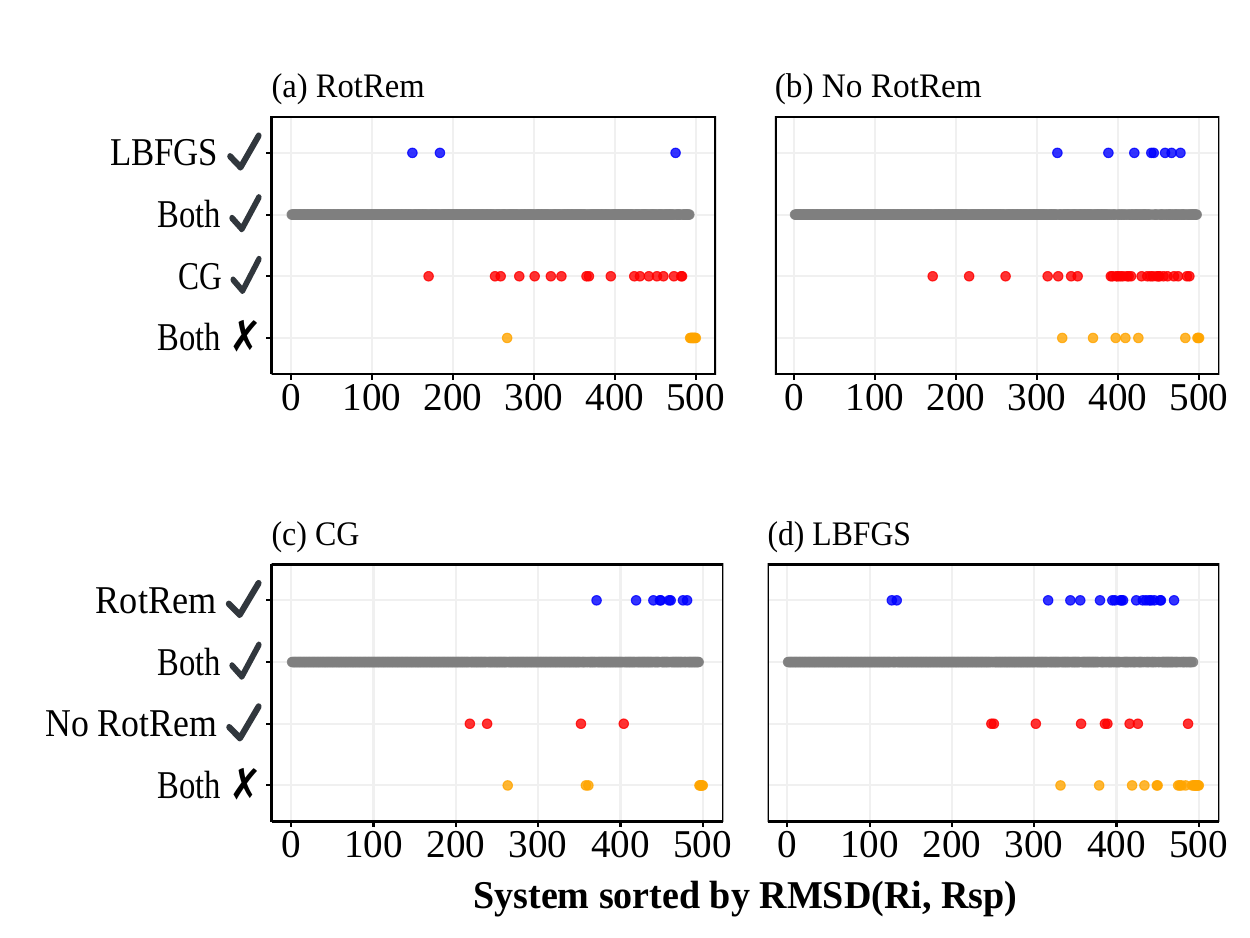}
\caption{\label{fig:successcompeda}Binary convergence outcomes for dimer method calculations across 500 systems, ordered by median root-mean-squared-displacement (RMSD) of the intial geometry relative to the saddle point geometry. Each panel directly compares two settings, with the vertical position and color indicating the outcome. Gray points indicate both compared settings succeed. Yellow points indicate where both compared settings fail. (a) Comparing LBFGS (blue) and CG (red) with rotation removal enabled, both methods (gray) succeeded in 94.0\% of systems, while CG additionally succeeded (red) in 3.6\% (LBFGS in 0.6\%, blue) and both failed (yellow) in 1.8\%. (b) Without rotation removal, comparing LBFGS (blue) to CG (red) yielded 90.8\% joint success (gray), with 5.8\% (red) unique CG successes (LBFGS in 1.6\%, blue) while the joint failure is 4.2\% (yellow). (c) Contrasting rotation removal (blue) with no rotation removal (red) using the CG-rotations shows high joint success (95.8\%, gray) with 0.8\% additional (red) without removal (1.8\% with rotation removal, blue) and 1.6\% cases where both fail (yellow). (d) For the LBFGS optimizer, comparing across rotation removal (blue) resulted in 90.6\% joint success (gray), without rotation removal (red) 1.8\% additional successes while with rotation removal (blue) 4\% additional cases succeed, and there are 3.6\% joint failures (yellow). Failures, particularly the unique failures with LBFGS without rotation removal, are concentrated at higher RMSD values.}
\end{figure}

Beyond the computational cost for successful runs, the overall convergence success is critical. Figure \ref{fig:successcompeda} presents a comparison of binary success outcomes across all 500 systems for which both methods in each comparison pair were run. High rates of joint success (where both compared methods converged) were generally observed across conditions. However, differences favouring the CG optimizer's robustness are apparent. When comparing optimizers directly, CG achieved unique success (succeeding when LBFGS failed) considerably more often than LBFGS did, both with rotation removal enabled (3.6\% vs 0.6\%, panel a) and disabled (5.8\% vs 1.6\%, panel b). Disabling rotation removal slightly reduced joint success rates and increased joint failures from 1.8\% to 4.2\% when comparing the optimizers. Examining the impact of rotation removal itself highlights CG's robustness further: using the CG optimizer (panel c), success rates remained very high (\textasciitilde{}96\% joint success) with minimal sensitivity, as disabling rotation removal resulted in unique success in only 0.8\% of cases. Conversely, using the LBFGS optimizer (panel d), disabling rotation removal significantly impacted performance; while joint success was still high (90.6\%), unique successes for the enabled rotation removal setting were more frequent (4.0\% vs 1.8\% for disabled), and the joint failure rate rose notably to 3.6\%. This suggests CG is less perturbed by the absence of rotation removal. As observed previously, convergence failures tend to concentrate at higher initial RMSD values, particularly in scenarios involving LBFGS without rotation removal.

This initial overview highlights apparent performance trends and differences between the method variants. However, these direct comparisons and visual inspections do not fully quantify the uncertainty associated with these effects or rigorously account for system-specific variability and potential interactions between the optimizer choice and rotation removal. Therefore, we employ the Bayesian hierarchical models detailed in Section \ref{sec:statmodels} for a more robust statistical analysis of these metrics.

It is important to note that there are alternative methods for removing global rotations and translations, such as those implemented in OPTIM \cite{walesEnergyLandscapesClusters2000}. In OPTIM, the Rayleigh-Ritz procedure \cite{munroDefectMigrationCrystalline1999} and subsequent partial tangent space minimization involve orthogonalization to overall translation or rotation. However, the application demonstrated here follows the quaternion-based method implemented in EON \cite{melanderRemovingExternalDegrees2015}.
\subsection{Effort analysis model}
\label{sec:comp_pes_model}
To quantify the effects of the dimer rotation optimizer and rotation removal on computational effort, we analyzed the number of PES calls using the Bayesian negative binomial generalized linear mixed-effects model described in Section \ref{sec:statmodels} (Eq. \ref{eq:brms_pes}). We fitted models corresponding to the different fixed-effects structures defined in Eq. \ref{eq:fixed_effects_structures}, examining main effects within relevant data subsets as well as the full interaction model using all comparable data.

\begin{equation}
\begin{aligned}
\text{PESCalls}_{ij} &\sim \text{NegativeBinomial}(\mu_{ij}, \phi) \\
\log(\mu_{ij}) &= \eta_{ij} \\
\eta_{ij} &=
    \begin{cases}
        \beta_0 + \beta_{1} \text{DR}_{i(j))} + u_j & \text{(a)} \\
        \beta_0 + \beta_{2} \text{RR}_{i(j))} + u_j & \text{(b)} \\
        \beta_0 + \beta_{1} \text{DR}_{i(j))} \\
        + \beta_{2} \text{RR}_{i(j))} + \beta_{3} (\text{DR}_{i(j))} \times \text{RR}_{i(j))}) + u_j & \text{(c)}
    \end{cases} \\
u_j &\sim \text{Normal}(0, \sigma^2_u)
\end{aligned}
\label{eq:model_pes_variants}
\end{equation}

First, applying the RotOptimizer model structure (Eq. \eqref{eq:model_pes_variants}a), specifically to calculations where rotation removal was disabled, we examin the effect of the rotation optimizer (\(\beta_{1}\)). Switching from the CG reference to the L-BFGS optimizer resulted in a small but statistically credible increase in the median number of PES calls by 2.6\% (95\% CrI: [1.1\%, 4.1\%]). This suggests a slight efficiency advantage, in terms of PES calls, for the CG optimizer under these conditions.

Next, using the RotRemoval model structure (Eq. \eqref{eq:model_pes_variants}b) applied to calculations using the CG optimizer, we examine the effect of enabling rotation removal (\(\beta_{2}\)). Enabling rotation removal led to a substantial and credible increase in the median PES calls by 41.5\% (95\% CrI: [38.6\%, 44.4\%]) compared to disabling it. This indicates a significant computational cost penalty associated with enabling rotation removal when using the CG optimizer in this setup.

\begin{figure}[htbp]
\centering
\includegraphics[width=.9\linewidth]{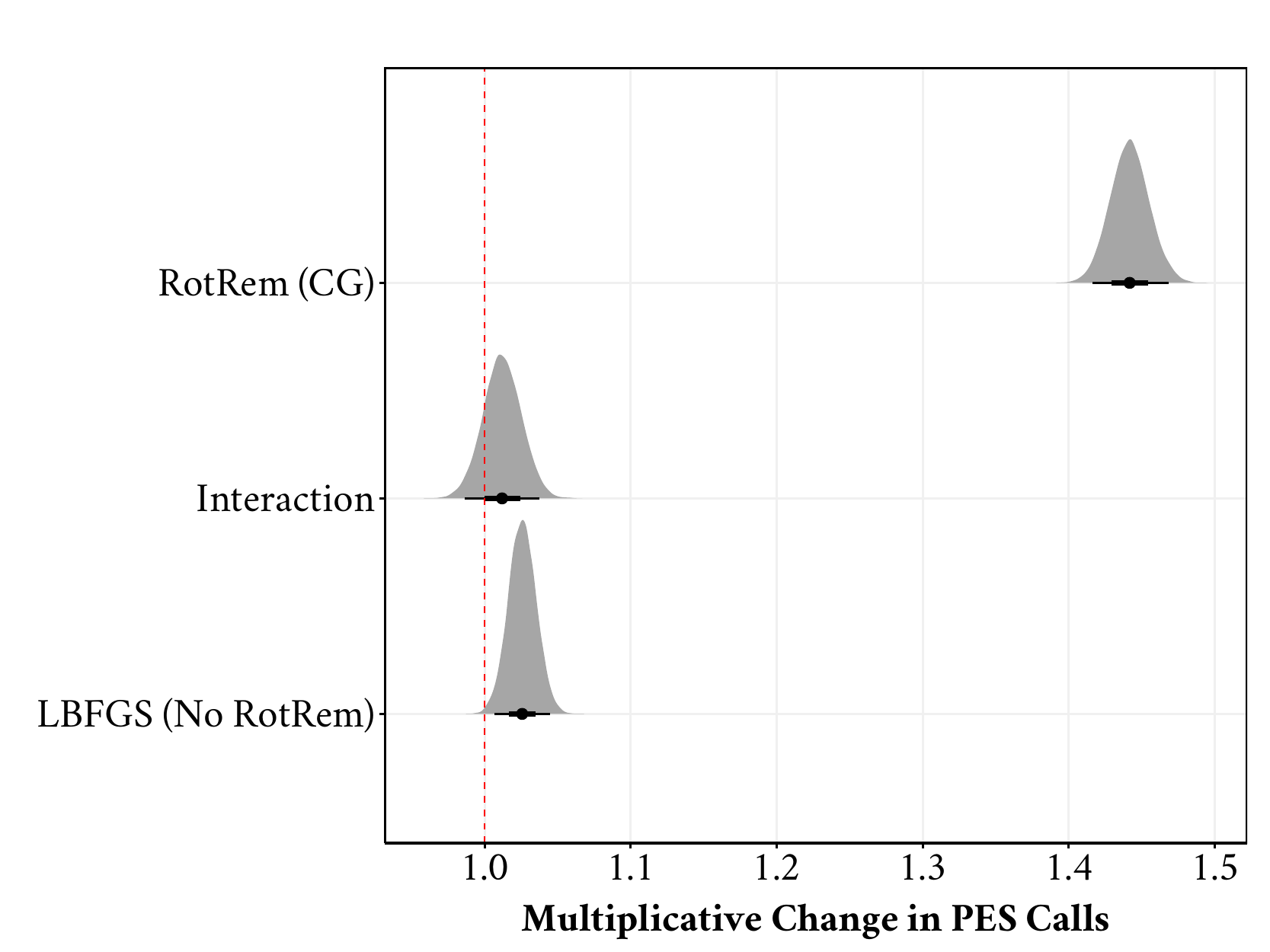}
\caption{\label{fig:effort_posteriors}Posterior distributions of fixed effects on the multiplicative scale for PES calls from the full interaction model. The x-axis represents the factor by which expected PES calls change relative to the baseline (CG optimizer, rotation removal disabled), with the dashed line at x=1 indicating no effect. Distributions show the main effect of using L-BFGS (vs CG, when rotation removal is disabled), the main effect of enabling rotation removal (vs disabling, when using the CG optimizer), and their interaction. Thick lines represent 95\% credible intervals with median dots. Enabling rotation removal shows a large multiplicative increase (\textasciitilde{}1.44), L-BFGS shows a small increase (\textasciitilde{}1.03), and the interaction credibly includes 1.}
\end{figure}

Finally, the full interaction model (Eq. \eqref{eq:model_pes_variants}c), incorporating both factors and their interaction simultaneously across 1805 valid observations, corroborated the main effects observed in the subset models (posterior distributions shown in Figure \ref{fig:effort_posteriors}). The estimated median increase in PES calls for L-BFGS relative to CG (when rotation removal is off, \(\beta_1\)) was 2.6\% (95\% CrI: [0.7\%, 4.5\%]), corresponding to a multiplicative factor credibly just above 1 (median 1.03). Similarly, the median increase for enabling rotation removal relative to disabling it (when using CG, \(\beta_{2}\)) was substantial at 44.2\% (95\% CrI: [41.6\%, 46.8\%]), clearly visualized by the distribution centered around a multiplicative factor of 1.44 in Figure \ref{fig:effort_posteriors}. The 95\% credible interval for the interaction term (\(\beta_{3}\)), visually overlapping 1 in Figure \ref{fig:effort_posteriors} (multiplicative factor median 1.01, 95\% CrI [0.99, 1.04]), provides no strong evidence that the effect of the optimizer choice significantly depends on the rotation removal setting, or vice versa. Furthermore, this full model estimated the standard deviation of the random intercepts for each system to be substantial (median \(\sigma_{u}\) = 0.63, 95\% CrI: [0.59, 0.67] on the log scale), indicating significant system-specific variation in the baseline number of PES calls required for convergence, independent of the method variations.

These findings from the hierarchical models, which account for system variability via random intercepts (\(\sigma_{u}\approx\) 0.63), reinforce the trends observed in the exploratory data analysis (Fig. \ref{fig:rotoptcomp} and Fig. \ref{fig:successcompeda}). They quantify the consistent, albeit small, PES call advantage of CG over L-BFGS (\(\beta_{1}\)) and the considerable computational overhead incurred by enabling this particular rotation removal implementation (\(\beta_2\)) for this dataset. The model diagnostics, including posterior predictive checks and residual analyses (shown for individual models in the preceding code sections, further details in SI), indicate reasonable model fit for the purpose of comparing these relative effects, despite some remaining heteroscedasticity common in complex chemical datasets.

\subsection{Success rate model}
\label{sec:org818a004}
We next analyzed the factors influencing the probability of successful convergence using the Bayesian Bernoulli generalized linear mixed-effects model specified in Section \ref{sec:statmodels} (Eq. \ref{eq:model_success_variants}). Effects are reported as Odds Ratios (OR), representing the multiplicative change in the odds of success compared to a reference condition. An OR > 1 indicates increased odds, OR < 1 indicates decreased odds, and a 95\% Credible Interval (CrI) that excludes 1 suggests a statistically credible effect.

\begin{equation}
\begin{aligned}
\text{Success}_{ij} &\sim \text{Bernoulli}(p_{ij}) \\
\text{logit}(p_{ij}) &= \log\left( \frac{p_{ij}}{1-p_{ij}} \right) = \eta_{ij} \\
\eta_{ij} &=
    \begin{cases}
        \beta_0 + \beta_{1} \text{DR}_i + u_j & \text{(a)} \\
        \beta_0 + \beta_{2} \text{RR}_i + u_j & \text{(b)} \\
        \beta_0 + \beta_{1} \text{DR}_i + \beta_{2} \text{RR}_i + \beta_{3} (\text{DR}_i \times \text{RR}_i) + u_j & \text{(c)}
    \end{cases} \\
u_j &\sim \text{Normal}(0, \sigma^2_u)
\end{aligned}
\label{eq:model_success_variants}
\end{equation}

First, applying the RotOptimizer model structure (Eq. \eqref{eq:model_success_variants}a) specifically to the subset of calculations where rotation removal was disabled, we examined the effect of the optimizer (\(\beta_{1}\)). L-BFGS showed significantly lower odds of success compared to the CG reference. The estimated median OR for L-BFGS relative to CG was 0.3 (95\% CrI: [0.13, 0.60]), indicating that the odds of convergence with L-BFGS were roughly only 30\% of the odds with CG under these conditions. This model also revealed substantial variability between systems in their baseline odds of success (posterior median \(s_u\) = 2.6 on the log-odds scale, 95\% CrI: [1.67, 3.81]).

Next, using the RotRemoval model structure (Eq. \eqref{eq:model_success_variants}b) specifically to the subset of calculations using CG-rotations, we examined the effect of rotation removal (\(\beta_{2}\)). The median OR for enabling rotation removal versus disabling it was 2.0 (95\% CrI: [0.73, 6.35]). Although the point estimate suggests doubled odds of success when enabling rotation removal, the wide credible interval comfortably includes 1. Consequently, this model provides no strong statistical evidence that enabling rotation removal significantly alters the odds of success compared to disabling it when using the CG optimizer. Substantial between-system variability in baseline success odds was again noted (median sd(Intercept) = 4.2, 95\% CrI: [2.63, 6.91] from this model's summary; the full model estimate is used below for consistency).

\begin{figure}[htbp]
\centering
\includegraphics[width=.9\linewidth]{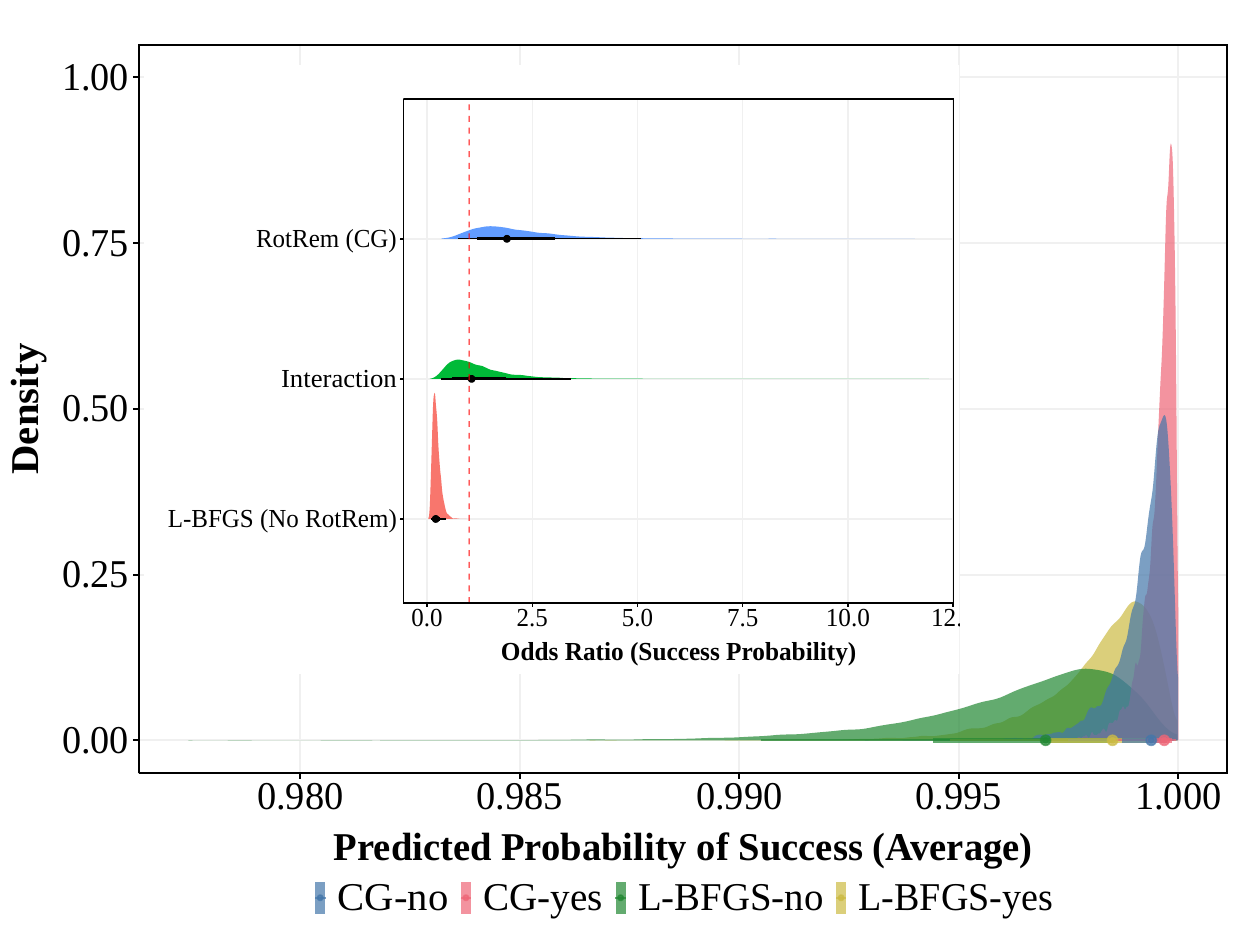}
\caption{\label{fig:success_posteriors}Posterior distributions for convergence success derived from the full interaction model (brms\textsubscript{pf}\textsubscript{idall}). The main panel displays the predicted probability of success for an average system under the four method conditions (Optimizer-Rotation Removal, see legend), illustrating high probabilities (>0.98) for all settings. The inset shows the posterior distributions for the fixed effects on the Odds Ratio (OR) scale relative to the baseline (CG, rotation removal disabled), with the dashed line at OR=1 indicating no change in odds. While absolute probabilities are high, the inset quantifies the relative effects, showing the OR for L-BFGS vs CG (bottom distribution) is credibly below 1, while the ORs for rotation removal (top) and the interaction (middle) credibly include 1. The distributions for CG are slightly shifted towards higher probabilities compared to L-BFGS. Distributions for enabling ('yes') versus disabling ('no') rotation removal are nearly identical within each optimizer column, but do tend to cause a rightward shift. Thick lines in the inset represent 95\% credible intervals with median dots.}
\end{figure}

Finally, the full interaction model (Eq. \eqref{eq:model_success_variants}c), analyzing all 2000 observations across the four conditions, allowed for a detailed assessment of success odds. While the predicted probability of success for an average system is very high (>0.98) under all conditions, as visualized in the main panel of Figure \ref{fig:success_posteriors}, the model revealed statistically credible differences when analyzing the underlying odds ratios as seen in the inset of Figure \ref{fig:success_posteriors}. It reinforced the substantially lower odds of success for L-BFGS compared to CG (median OR = 0.2, 95\% CrI: [0.09, 0.45]); this relative difference, indicating higher robustness for CG, is subtly reflected in Figure \ref{fig:success_posteriors} by the slight shift of the L-BFGS probability distributions (right column) towards lower values compared to CG (left column). The model also showed no credible main effect for rotation removal (median OR = 1.9, 95\% CrI: [0.74, 5.07]) nor a credible interaction (median Interaction OR = 1.1, 95\% CrI: [0.33, 3.42]), consistent with the high degree of similarity between the top ('no') and bottom ('yes') panels within each optimizer in Figure \ref{fig:success_posteriors}. Thus, while both optimizers perform well on average, the primary factor credibly influencing the relative odds of success in this analysis is the choice of optimizer, with CG demonstrating significantly higher odds of convergence than L-BFGS across both rotation removal settings in this dataset. The large estimated standard deviation for the system-specific random intercepts from the full model (posterior median \(\sigma_u\) = 3.6, 95\% CrI: [2.79, 4.80]) underscores that system-specific properties remain a major determinant of convergence success or failure for individual cases, explaining deviations from the high average success probability. Model diagnostics indicated good convergence and fit (details in SI).

The preceding analyses provide a quantitative, statistically robust assessment of how the choice of dimer rotation optimizer and the application of rotation removal influence both the computational effort (PES calls) and the likelihood of success for saddle point searches within this benchmark set. Having established these effects and accounted for system-specific variability using Bayesian hierarchical models, we now turn to discussing the implications of these findings in the context of theoretical expectations, practical algorithm choice, and the broader methodology of benchmarking computational chemistry methods.
\section{Discussion}
\label{sec:discussion}
Our analysis, employing Bayesian hierarchical models to account for system-specific variability, provides quantitative insights into the performance of CG and L-BFGS optimizers for dimer rotation, and the effect of enabling rotation removal, across a large benchmark set. The key findings indicate a small but credible PES call advantage and significantly higher convergence odds for the CG optimizer compared to L-BFGS. Conversely, enabling rotation removal incurred a substantial PES call penalty without providing a statistically credible improvement in success probability in the full model applied to this dataset.

The superior performance of CG over L-BFGS, particularly its enhanced robustness demonstrated by the success probability analysis (OR for L-BFGS vs CG \(\approx\) 0.2 [0.09, 0.45]), aligns with potential theoretical advantages of CG in navigating complex potential energy surfaces. While modern L-BFGS methods incorporate strategies to ensure positive-definite Hessian approximations \cite{nocedalNumericalOptimization2006}, they may still be susceptible to challenges in regions with complex curvature compared to methods like CG that directly seek movement along specific eigenmodes \cite{lengEfficientSoftestMode2013}. The observed difference in computational effort per successful run was modest (L-BFGS requiring \textasciitilde{}2.6\% [0.7\%, 4.5\%] more PES calls than CG when rotation removal was off), suggesting that the primary practical advantage of CG in this context lies in its higher probability of reaching convergence.

Counter-intuitively, enabling rotation removal, theoretically designed to simplify the optimization by removing global rotations of the coordinate vector  \cite{melanderRemovingExternalDegrees2015}, resulted in a significant increase in computational cost (median +44\% PES calls vs disabled) without a corresponding statistically credible improvement in success probability in our full model (OR for RotRem Yes vs No \(\approx\) 1.9 [0.74, 5.07]). This discrepancy likely highlights the sensitivity of optimization performance to implementation details and the specific nature of the PES. The projection involved in rotation removal interacts with the optimizer's steps and Hessian updates; for this dataset containing potentially fragmenting or aggregating systems near saddle points, where the separation of internal and external coordinates can be ill-defined \cite{nicholsWalkingPotentialEnergy1990}, this projection may hinder convergence pathways that are accessible without it. Thus we observe less efficient trajectories when rotation removal is active for the majority of these systems.

We underscore the value of applying rigorous statistical modeling to computational benchmarks, moving beyond simple averages or qualitative comparisons. Traditional approaches can be easily confounded by the large system-to-system variability inherent in chemical datasets, as evidenced by the substantial standard deviation estimated for the random intercepts in our models i.e. \(\sigma_{u}\approx\) 0.63 for PES calls (on the logarithmic scale), \$\(\sigma\)\textsubscript{u}\(\approx\)\$3.6 for success odds(on the logit scale). Bayesian hierarchical models explicitly account for this variability through random effects (\(u_j\)). Furthermore, they leverage partial pooling (shrinkage), effectively borrowing strength across systems to produce more stable and reliable estimates for the effects of different methods, mitigating the influence of outlier systems - an effect sometimes colloquially related to ``regression towards the mean''. This framework also provides direct quantification of uncertainty via full posterior distributions and credible intervals (e.g., Figs. \ref{fig:effort_posteriors}, \ref{fig:success_posteriors}), facilitating more nuanced conclusions than point estimates or p-values alone \cite{mcelreathStatisticalRethinkingBayesian2020,mcshaneAbandonStatisticalSignificance2019}. The ability to model interactions formally is another advantage for dissecting performance factors.

Further theoretical considerations help to contextualize these empirical results. The rotational phase of the Dimer method represents a specialized optimization problem: isolating a single minimum-curvature mode. The Conjugate Gradient algorithm is inherently well-suited for such eigenvector-following tasks. It can also be less susceptible to being trapped by higher-energy eigenpairs during the search \cite{lengEfficientSoftestMode2013}. Conversely, while L-BFGS is a highly effective general-purpose optimizer, it does not, in most cases, share the same quadratic convergence guarantees as the full BFGS method \cite{berahasLimitedmemoryBFGSDisplacement2022}. This theoretical background aligns well with our findings, where the most significant performance gap between the two methods appeared in their relative robustness and success rates, rather than in the computational cost of already successful runs.

Given the results, particularly the higher robustness of CG and the general inefficiency introduced by enabling rotation removal in this context, a practical strategy for high-throughput workflows might involve defaulting to the CG optimizer without rotation removal. A fallback mechanism could then activate rotation removal only for specific systems that initially fail to converge, potentially capturing benefits for problematic cases where external modes interfere, although identifying such cases a priori remains challenging. This ``chain of methods'' approach warrants further study but illustrates how statistically grounded benchmarks can inform practical workflow design beyond simply selecting a single ``best'' method.

While this study has focused on variants of the Dimer method, the Bayesian hierarchical framework presented here is broadly applicable. An important future direction is its application to a comparative benchmark of fundamentally different classes of algorithms, such as single-ended methods (Dimer, Sella) versus chain-of-states methods (NEB, GSM). Such an analysis could quantitatively address long-standing questions, for example, by modeling the probability of finding the correct path as a function of the number of images in an NEB calculation or comparing the raw computational cost versus the likelihood of success for finding a transition state starting from only the reactant and product states. This would provide the community with a robust, data-driven basis for selecting the optimal algorithm class for a given high-throughput workflow.
\section{Conclusions}
\label{sec:concl}
We introduced and applied a Bayesian hierarchical modeling framework to
rigorously analyze performance metrics from computational chemistry benchmarks,
demonstrating its utility for comparing dimer method variants across a large,
diverse dataset. This statistically robust approach moves beyond simple averages
or visualizations, accounting for system-specific variability and providing
nuanced uncertainty quantification, offering a more actionable view for
benchmarking algorithms.

Our analysis of the Dimer method variants revealed that for this large benchmark
set, the Conjugate Gradient (CG) optimizer for dimer rotation provided a clear
advantage in convergence robustness over L-BFGS (Section \ref{sec:statmodels}).
The decision to enable rotation removal is more complex. While theoretically
motivated to simplify the potential energy surface, our findings indicate it led
to higher computational costs for the majority of systems studied. However, our
models also suggest that rotation removal may offer a reliability benefit when
paired with the L-BFGS optimizer, highlighting a subtle interplay between
algorithmic components that warrants further study.

We assert that these results, along with the automated capabilities of most
workflow engines suggest a ``chain'' of methods approach rather than an absolute
best. Although given the slight increase in performance and the significant
increase in robustness it is hard to argue for the LBFGS rotations, the
situation for the chemically intuitive external degree of freedom removal is
more nuanced.

From the generalized linear mixed models (Section \ref{sec:statmodels}), using the
LBFGS for rotations showed a small but statistically discernible increase of
3.4\% (95\% CrI: [1.1\%, 5.8\%]) in PES calls. Furthermore, the Bernoulli model
suggests a substantial improvement in success probability for Dimer CG, with an
estimated 23\% higher success rate (95\% CI: 5\%, 40\%), supporting the conclusion
that CG-rotations are more robust.

There are however, situations where the rotation removal is required
for convergence. Since the data demonstrates that for these random saddle point
initializations that the rotation removal for such fragmenting systems is not
optimal in most cases, it is more natural to first run CG-rotation without
external rotation removal, and have a fallback rule to run CG-rotation with
external rotation only for systems which otherwise fail. This strategy leverages
the higher success rate of CG in the majority of cases while providing a
fallback option for more challenging systems.

Our benchmark results and analysis are packaged in a user-friendly repository,
showing integration with Snakemake for high throughput calculations and EON for
the energy and forces calculation through NWChem along with \texttt{R} code for the
figures and running the Bayesian analysis. It is also, to the best of our
knowledge, the largest benchmark study done on the modalities of the dimer
method. Our analysis provides a more nuanced and statistically
robust assessment of performance than traditional approaches, which often rely
on simple averages or visualizations that can obscure important details.
We expect this reproducible resource will be indispensible to others in
the community performing high throughput chemistry or algorithm development.
\section{Supplementary Material}
\label{sec:orgd30338c}
Model validations for the Bayesian models are fairly standard and verbose, so
they are present in the accompanying supplementary material.
\section*{Acknowledgments}
\label{sec:org5686074}
This work was supported by the Icelandic Research Fund (grant no. 217436-053).

RG thanks Dr. Miha Gunde, Dr. Amrita Goswami, Dr. Moritz Sallermann, Prof.
Debabrata Goswami, Prof. Laurent Karim Béland, Prof. Morris Riedel, Mrs. Sonaly
Goswami and Mrs. Ruhila Goswami for discussions. RG also thanks Prof. Normand
Mousseau and Prof. Birgir Hrafnkelsson for manuscript suggestions and edits.
\section*{Data Availability Statement}
\label{data-availability-statement}
The full computational workflow, analysis code, and data supporting this study are openly available. A version-tagged GitHub repository (\url{https://github.com/HaoZeke/brms_idrot_repro}) provides all resources needed to reproduce the findings. To ensure full reproducibility, the repository contains a complete computational environment managed by pixi, with version-pinned software dependencies including EON, NWChem, and ASE, alongside utilities from Wailord \cite{goswamiWailordParsersReproducibility2022}. In line with best practices for literate and reproducible programming \cite{goswamiHighThroughputReproducible2022,goswamiReproducibleHighPerformance2022,sallermannFlowyHighPerformance2024,goswamiEfficientImplementationGaussian2025}, some dependencies are bundled directly via \texttt{gitsubrepo}. For permanent access and citation, the final curated dataset is hosted on the Materials Cloud Archive \cite{talirzMaterialsCloudPlatform2020} under the entry \citet{goswamiDatasetBayesianHierarchical2025}.
\section*{Conflict of interest}
\label{sec:org5dd2e82}
The authors declare no conflict of interest.

\begin{appendices}
\vspace{0.7em}
\noindent \textbf{Exact code to reproduce all models is part of the repository.}
\section{Statistical model validation}
\label{sec:org073f0ae}
For details of the specification and interpretation refer to the main text. To ensure the supplementary can be read independently, recall the fixed effect structures i.e.
\begin{equation}
\text{Fixed Effects}_{ij} =
    \begin{cases}
        \beta_0 + \beta_{1} \text{DR}_{i(j))} & \text{(RotOptimizer)} \\
        \beta_0 + \beta_{2} \text{RR}_{i(j))} & \text{(RotRemoval)} \\
        \beta_0 + \beta_{1} \text{DR}_{i(j))} + \beta_{2} \text{RR}_{i(j))} + \beta_{3} (\text{DR}_{i(j))} \times \text{RR}_{i(j))}) & \text{(Full)}
    \end{cases}
\label{eq:fixed_effects_structures}
\end{equation}
For the analysis below, log-scale parameters mean the direct estimates of the CI from the model summary cannot be used, and instead simulated draws from the posterior are individually exponentiated and then averaged to obtain the point estimates. Details and code for this proceedure are in the accompanying Github repository.
\subsection{Performance models}
\label{sec:orga5b946d}
\subsubsection{Optimizer for the rotation phase}
\label{sec:org6ab9e6f}
Here we use the (RotOptimizer) subset, i.e. with no rotation removal.

\begin{verbatim}
 Family: negbinomial
  Links: mu = log; shape = identity
Formula: pes_calls ~ dimer_rot + (1 | mol_id:spin)
   Data: data (Number of observations: 896)
  Draws: 4 chains, each with iter = 4000; warmup = 1000; thin = 1;
         total post-warmup draws = 12000

Multilevel Hyperparameters:
~mol_id:spin (Number of levels: 456)
              Estimate Est.Error l-95%
sd(Intercept)     0.62      0.02     0.58     0.67 1.01      555     1169

Regression Coefficients:
               Estimate Est.Error l-95%
Intercept          5.69      0.03     5.64     5.75 1.02      286      593
dimer_rotlbfgs     0.03      0.01     0.01     0.04 1.00    24427     8675

Further Distributional Parameters:
      Estimate Est.Error l-95%
shape   129.21     11.49   108.17   153.03 1.00     6860     8842

Draws were sampled using sample(hmc). For each parameter, Bulk_ESS
and Tail_ESS are effective sample size measures, and Rhat is the potential
scale reduction factor on split chains (at convergence, Rhat = 1).
\end{verbatim}

The analysis revealed a small, but statistically discernible, difference between
the CG and LBFGS algorithms. The estimated coefficient for \texttt{dimer\_rotlbfgs} was
0.03 (95\% CI: 0.01, 0.04).  This indicates that the LBFGS method
is associated with a slightly higher number of PES calls compared to the CG
method.

The random intercept standard deviation (0.62, 95\% CI: 0.58, 0.67) is
considerably large, suggesting more substantial variability in baseline
pes\textsubscript{calls} across molecule/spin combinations for the IDimer method. The shape
parameter of the negative binomial distribution was estimated to be 129.21 (95\%
CI: 108.17, 153.03), indicating low overdispersion.

\begin{figure}[htbp]
\centering
\includegraphics[scale=0.8]{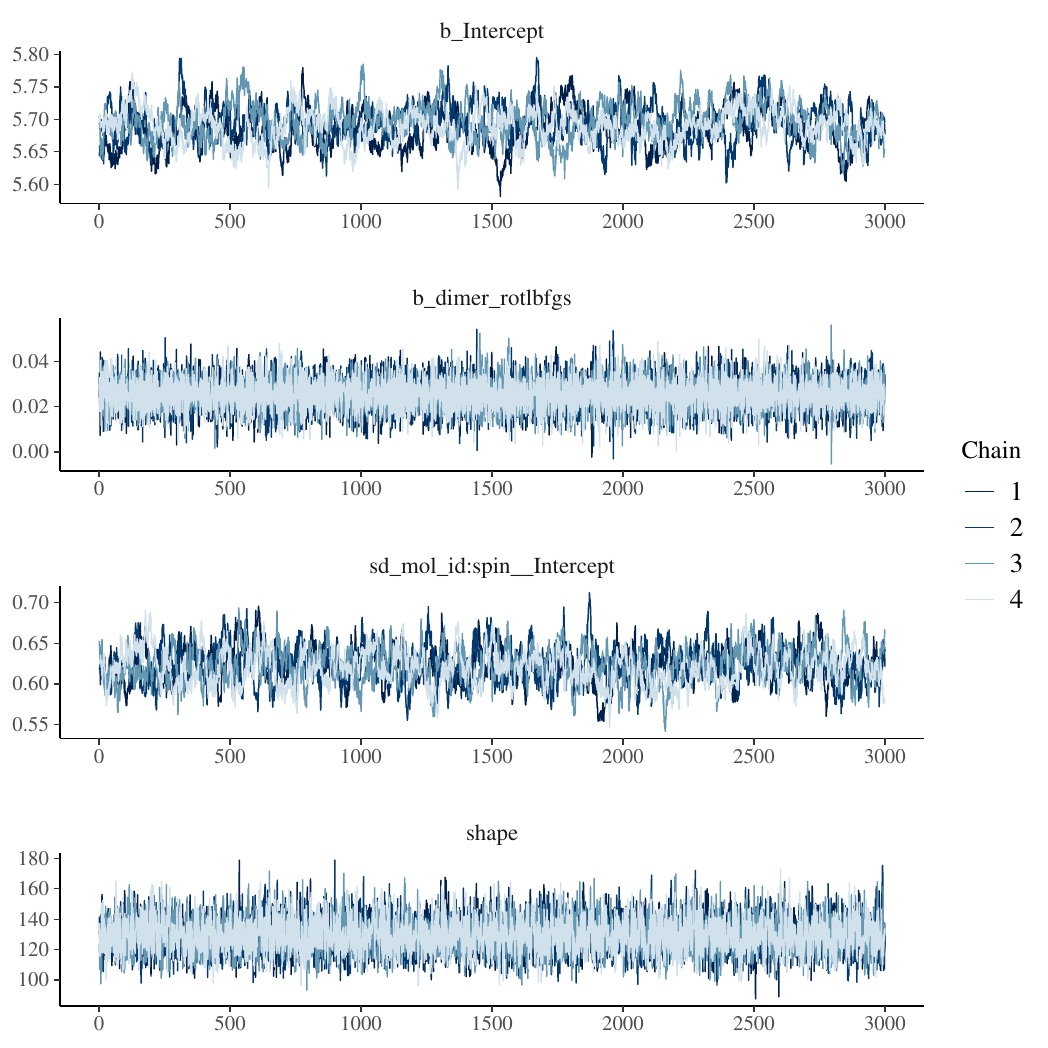}
\caption{\label{fig:brms_pes_idcg_idlbfgs_trace}Trace plots for the IDimer CG vs. LBFGS comparison, showing well-mixed chains.}
\end{figure}

Chain convergence was excellent (all \texttt{Rhat} values = 1.00, and all \texttt{Bulk\_ESS} and \texttt{Tail\_ESS} values were large). Trace plots (Figure \ref{fig:brms_pes_idcg_idlbfgs_trace}) confirmed good mixing of the chains.

\begin{figure}[htbp]
\centering
\includegraphics[scale=0.5]{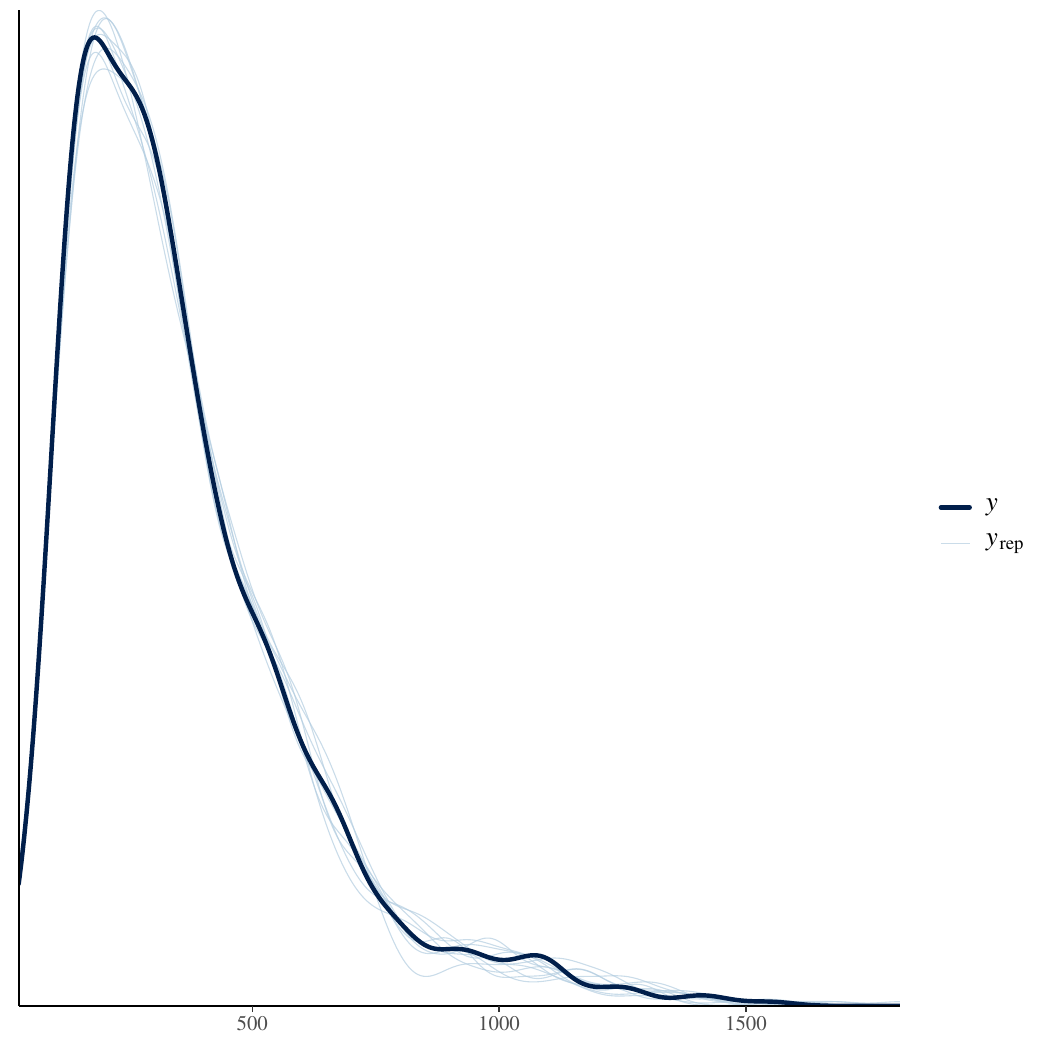}
\caption{\label{fig:brmsp_idcg_idlbfgs_pp}Posterior predictive check for the IDimer CG vs. LBFGS comparison (continuous).}
\end{figure}

Posterior predictive checks (Figure \ref{fig:brmsp_idcg_idlbfgs_pp}) indicate a good model fit, with the posterior predictive distribution closely matching the observed distribution of \texttt{pes\_calls}.

\begin{figure}[htbp]
\centering
\includegraphics[scale=0.5]{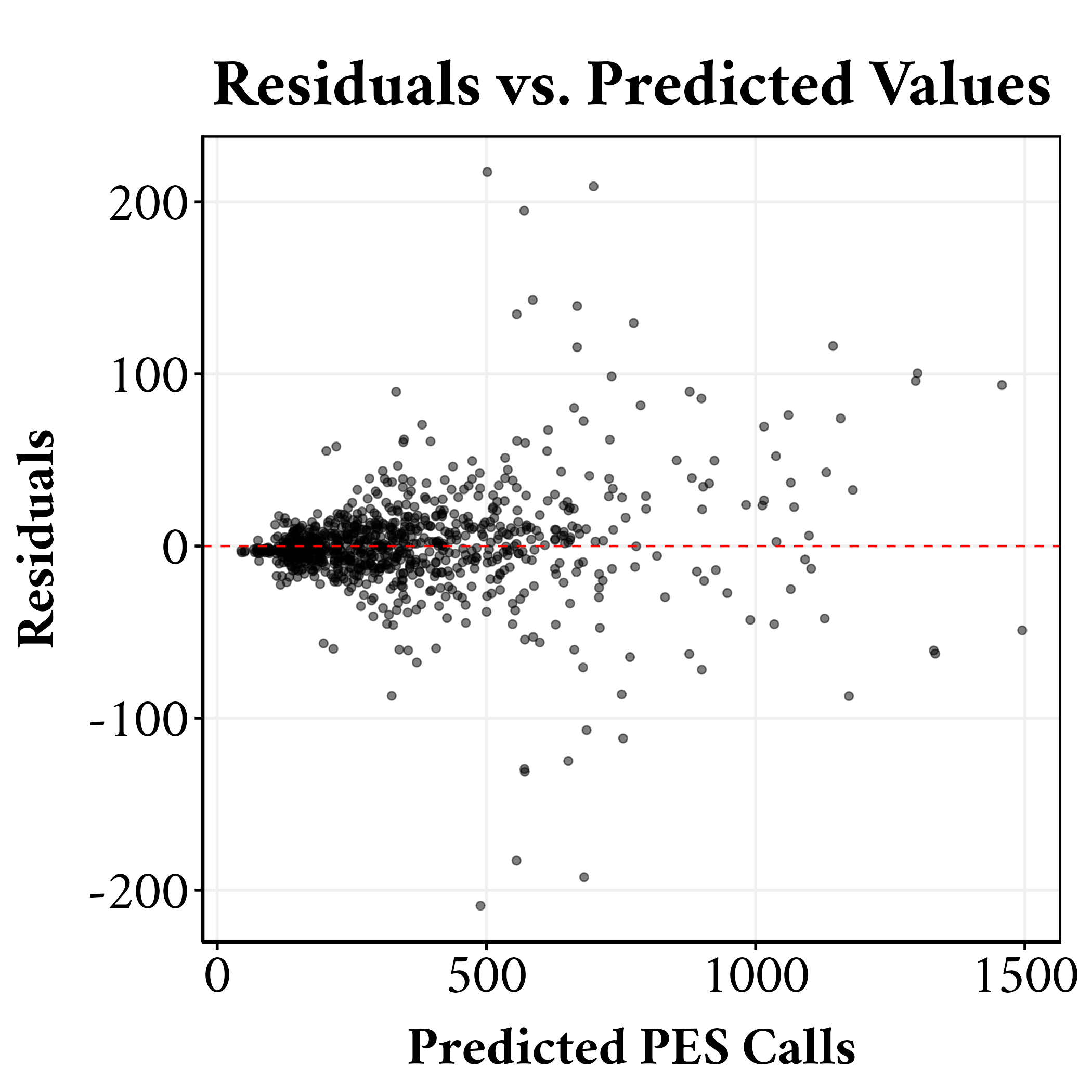}
\caption{\label{fig:brms_pes_clbfgs_norot_resid}Error residuals for the RotOptimizer subset performance model.}
\end{figure}

The residual error analysis in Figure \ref{fig:brms_pes_clbfgs_norot_resid} shows slight heteroscedasticity.

\begin{figure}[htbp]
\centering
\includegraphics[scale=0.5]{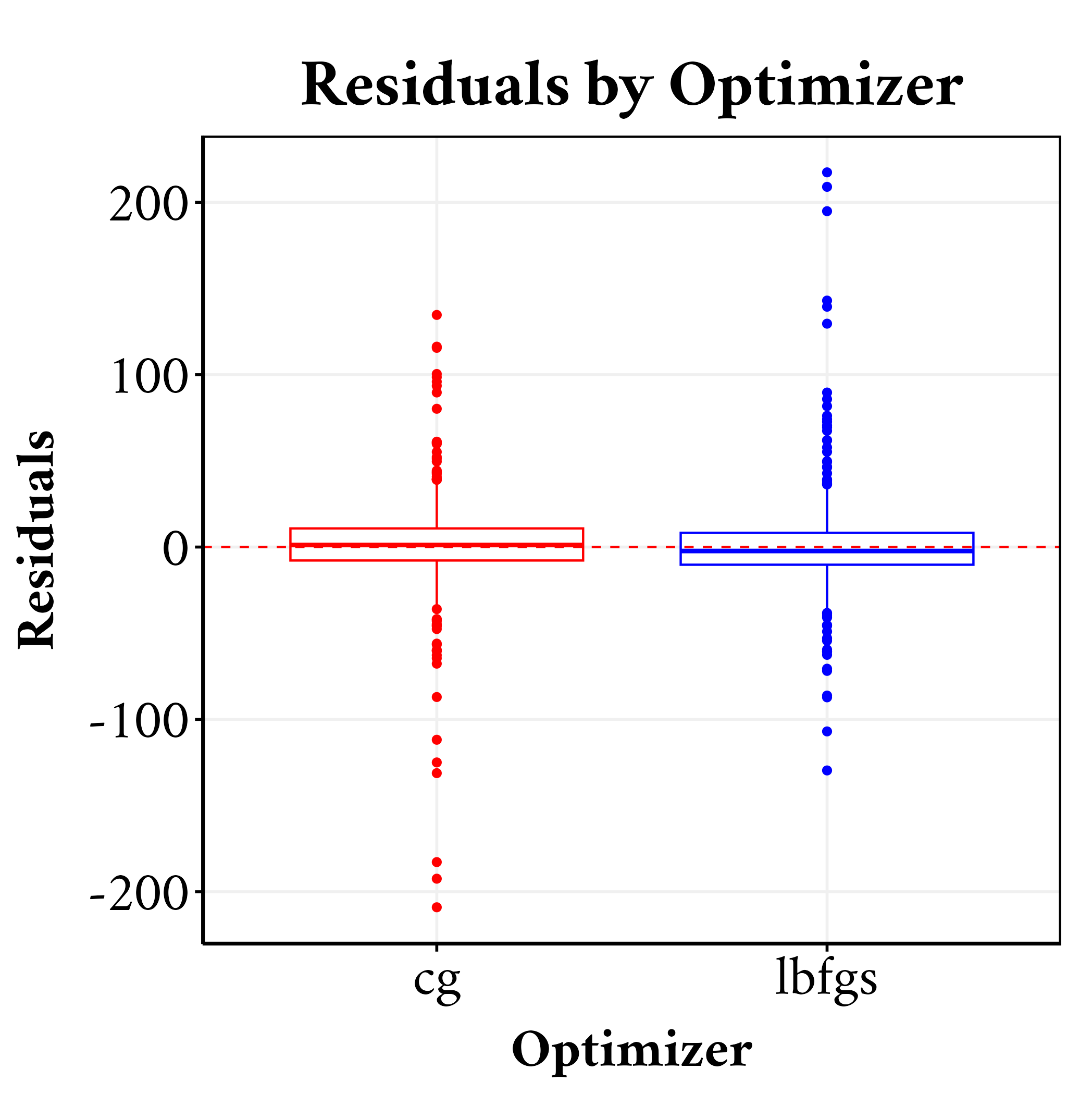}
\caption{\label{fig:brmsp_clbfgs_norot_boxplt}Boxplot of the error residuals for the RotOptimizer subset performance model, showing no bias.}
\end{figure}

Crucially, the box plot in Figure \ref{fig:brmsp_clbfgs_norot_boxplt} indicates that there is no systematic bias, so it is good enough for relative comparisons.

The posterior effects including the random effects are summarized in Table \ref{rotopt_perf}.

\begin{table}[htbp]
\caption{\label{rotopt_perf}RotOptimizer subset performance model effects showing credible interavals}
\centering
\begin{tabular}{lrl}
Effect Type & Median Effect & 95\% CrI\\
\hline
Expected PES Calls (CG, No RotRem) & 296.30 & {[}280.5, 313.5]\\
Multiplicative Factor (L-BFGS vs CG) & 1.03 & {[}1.01, 1.04]\\
Percentage Change (L-BFGS vs CG) & 2.6\% & {[}1.1\%, 4.1\%]\\
sd(Intercept) [molid:spin] & 0.62 & {[}0.58, 0.67]\\
\end{tabular}
\end{table}
\subsubsection{Removal of external rotation}
\label{sec:org6abba76}
This model is fit to the RotRemoval subset, with the CG optimizer, as it is the better of the two from the previous section.

\begin{verbatim}
 Family: negbinomial
  Links: mu = log; shape = identity
Formula: pes_calls ~ rot_removal + (1 | mol_id:spin)
   Data: data (Number of observations: 971)
  Draws: 4 chains, each with iter = 4000; warmup = 1000; thin = 1;
         total post-warmup draws = 12000

Multilevel Hyperparameters:
~mol_id:spin (Number of levels: 492)
              Estimate Est.Error l-95%
sd(Intercept)     0.66      0.02     0.62     0.71 1.01     1253     1380

Regression Coefficients:
               Estimate Est.Error l-95%
Intercept          5.76      0.03     5.70     5.82 1.01      695     1593
rot_removalyes     0.35      0.01     0.33     0.37 1.00    25842     8962

Further Distributional Parameters:
      Estimate Est.Error l-95%
shape    44.35      3.06    38.58    50.53 1.00     6202     9190

Draws were sampled using sample(hmc). For each parameter, Bulk_ESS
and Tail_ESS are effective sample size measures, and Rhat is the potential
scale reduction factor on split chains (at convergence, Rhat = 1).
\end{verbatim}

The analysis revealed a large, statistically discernible, difference between
choosing to use external rotation removal. The estimated coefficient for \texttt{rot\_removalyes}
was 0.35 (95\% CI: 0.33, 0.37). This indicates that the removal of
external rotations is associated with a significantly higher number of PES calls
compared to not using the rotation removal.

The random intercept standard deviation (0.66, 95\% CI: 0.62, 0.71) is
considerably large, suggesting more substantial variability in baseline
pes\textsubscript{calls} across molecule/spin combinations for the dimer method. The shape
parameter of the negative binomial distribution was estimated to be 44 (95\%
CI: 38.58, 50.53), indicating high overdispersion.

\begin{figure}[htbp]
\centering
\includegraphics[scale=0.8]{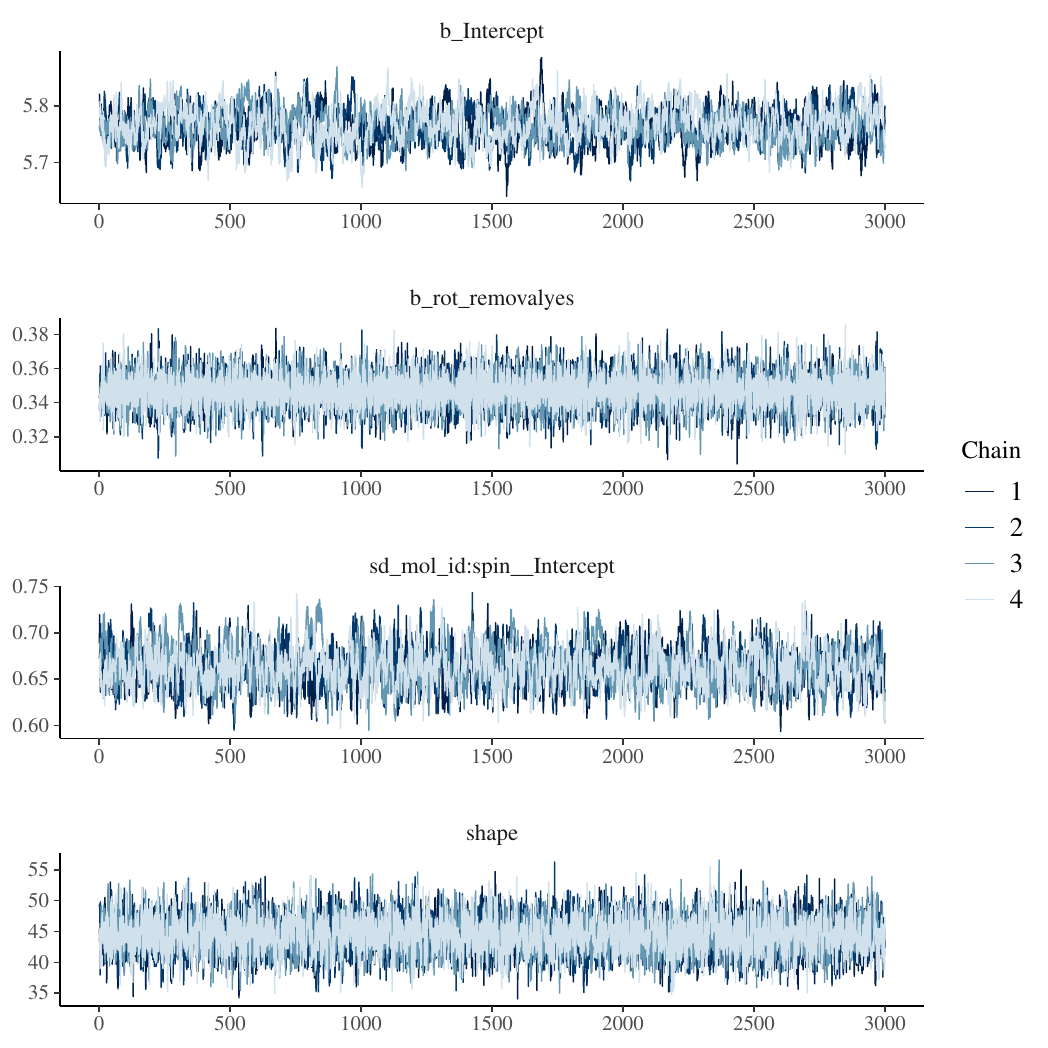}
\caption{\label{fig:brmsp_rotrem_traceplots}Trace plots for the rotation removal comparison, showing well-mixed chains.}
\end{figure}

Chain convergence was excellent (all Rhat values \(\sim\) 1.00, and all Bulk\textsubscript{ESS} and
Tail\textsubscript{ESS} values were large). Trace plots (Figure \ref{fig:brmsp_rotrem_traceplots})
confirmed good mixing of the chains.

\begin{figure}[htbp]
\centering
\includegraphics[scale=0.5]{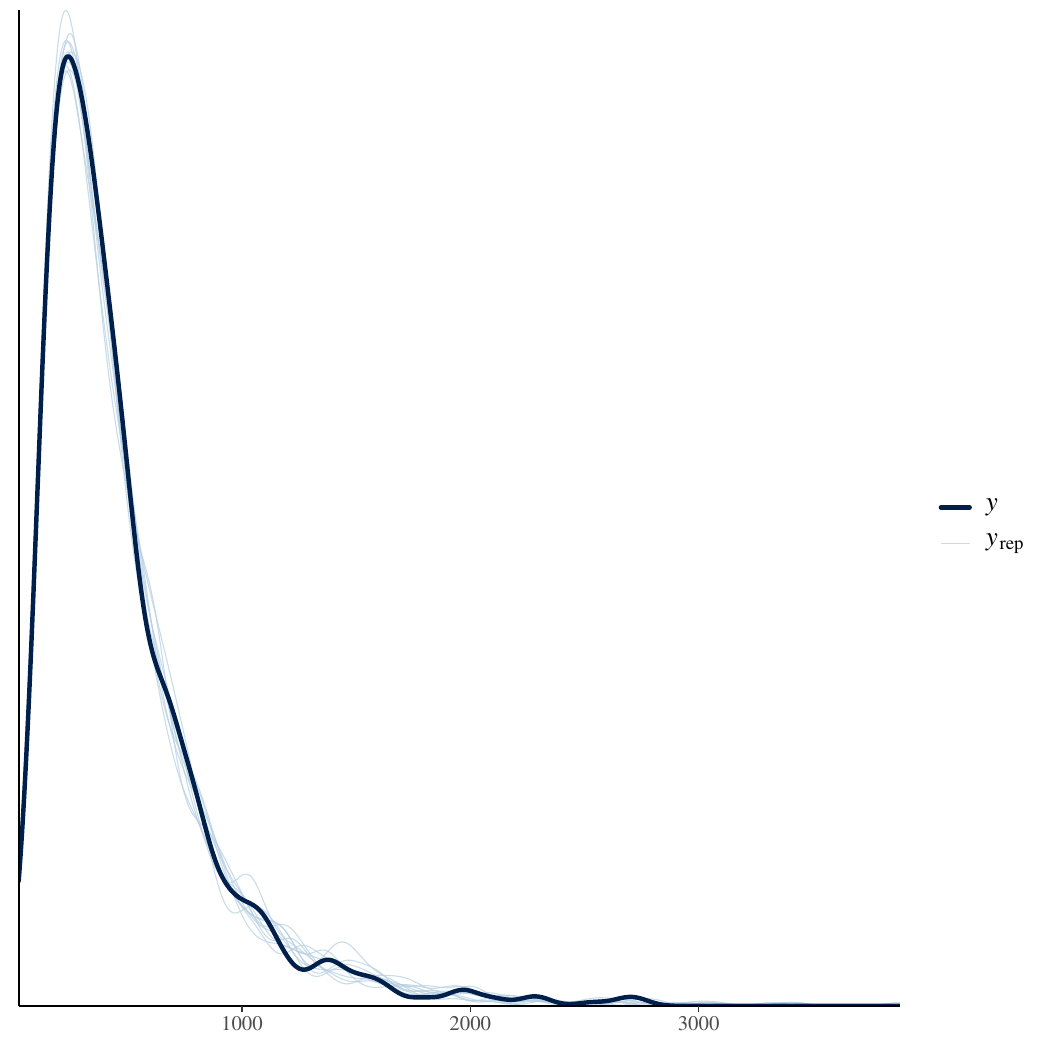}
\caption{\label{fig:brmsp_rotrem_pp}Posterior predictive check for the rotation removal comparison (continuous).}
\end{figure}

Posterior predictive checks (Figure \ref{fig:brmsp_rotrem_pp}) indicate a good model fit, with the posterior predictive distribution closely matching the observed distribution of \texttt{pes\_calls}.

\begin{figure}[htbp]
\centering
\includegraphics[scale=0.5]{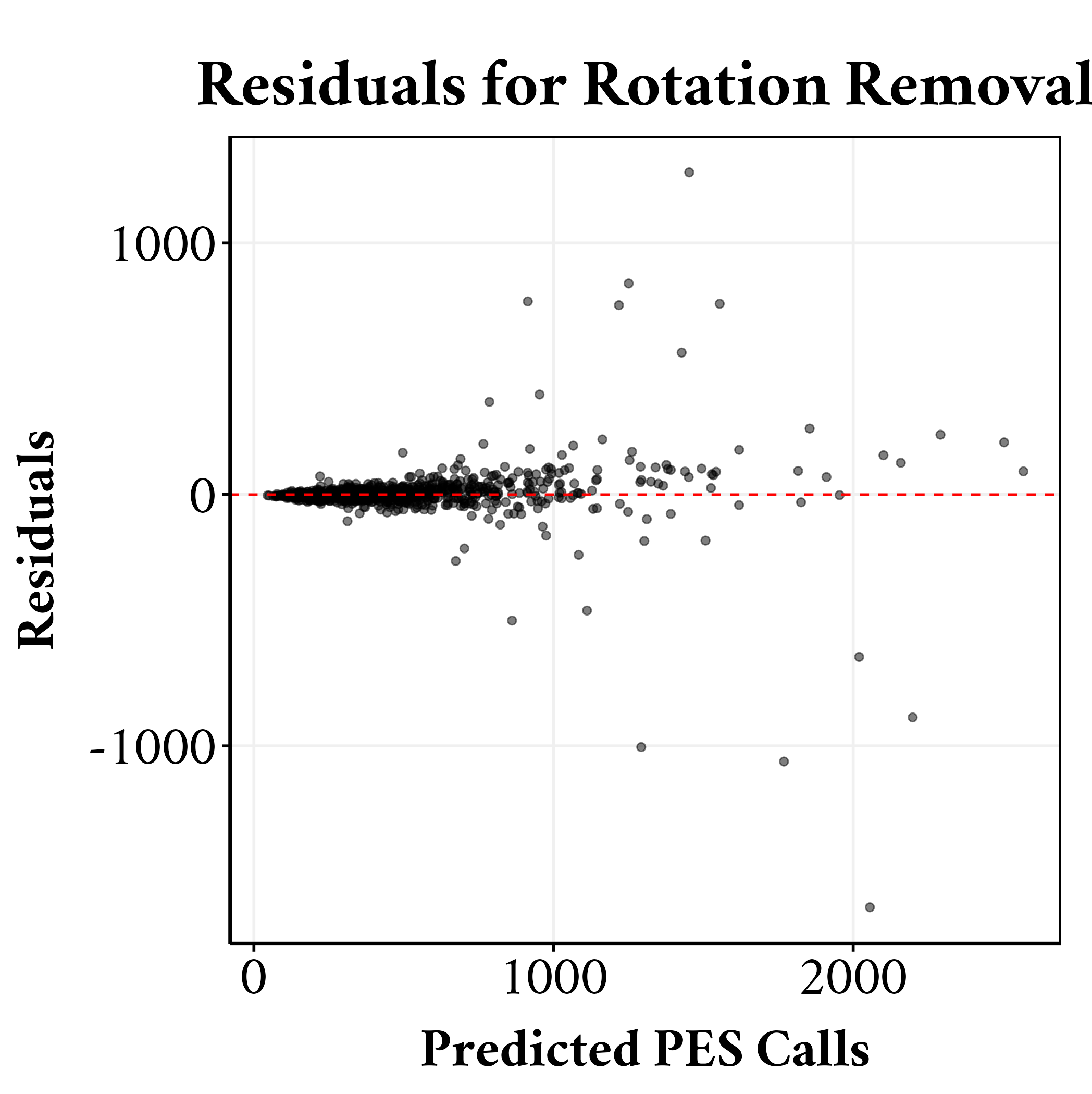}
\caption{\label{fig:brmsp_rotrem_resid}Error residuals for the RotRemoval subset performance model.}
\end{figure}

The residual error analysis in Figure \ref{fig:brmsp_rotrem_resid} shows slight heteroscedasticity, but less than the optimizer subset.

\begin{figure}[htbp]
\centering
\includegraphics[scale=0.5]{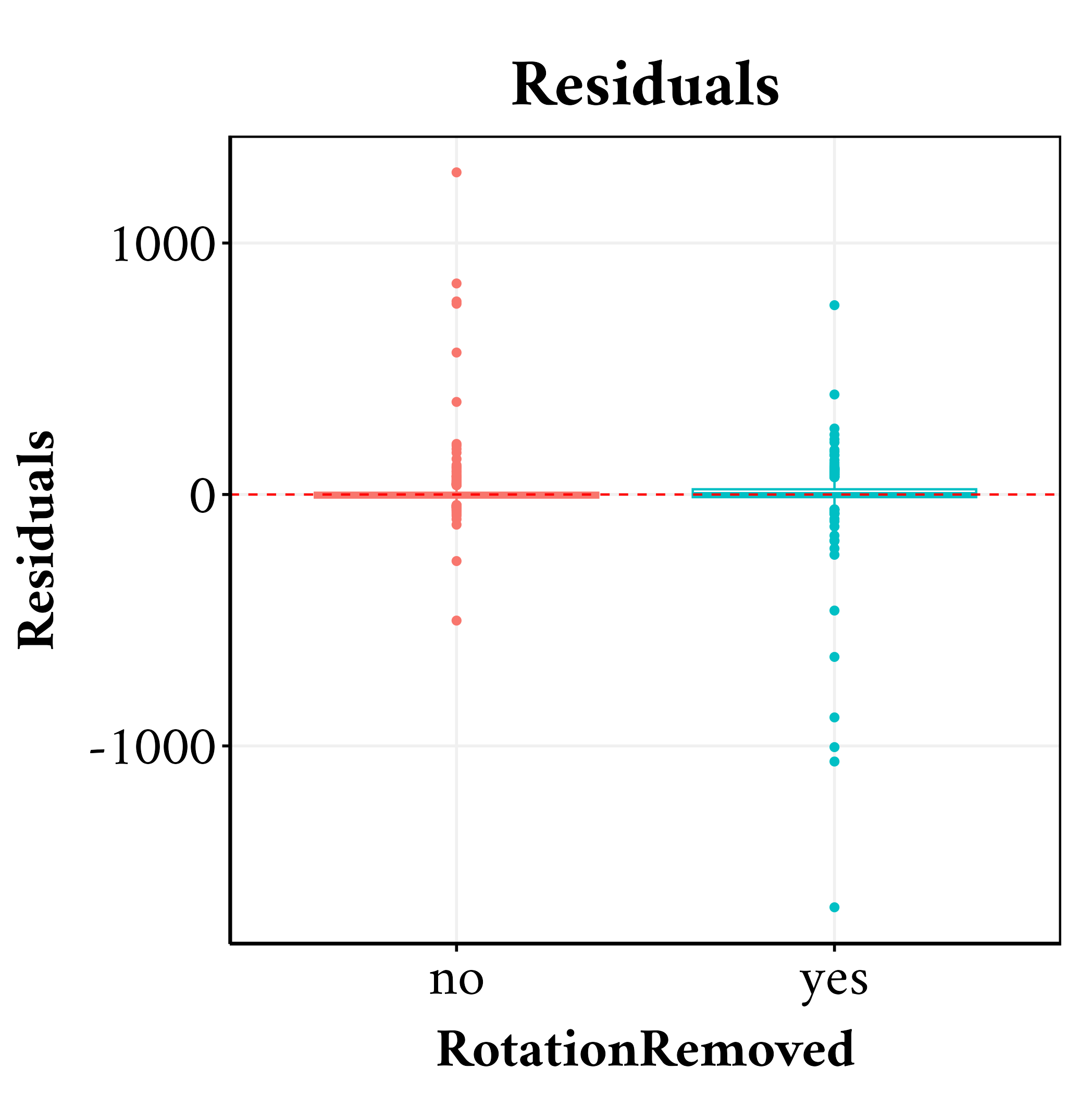}
\caption{\label{fig:brmsp_rotrem_boxplt}Boxplot of the error residuals for the RotRem subset performance model, showing no bias.}
\end{figure}

As before, the box plot in \ref{fig:brmsp_rotrem_boxplt} indicates that there is no systematic bias, so it is good enough for relative comparisons.

The posterior effects including the random effects are summarized in Table \ref{rotrem_perf}.

\begin{table}[htbp]
\caption{\label{rotrem_perf}RotRemoval subset performance model effects, including credible intervals.}
\centering
\begin{tabular}{lrl}
Effect Type & Median Effect & 95\% CrI\\
\hline
Expected PES Calls (CG, No RotRem) & 318.70 & {[}300.3, 338.1]\\
Multiplicative Factor (RotRem vs No RotRem) & 1.41 & {[}1.39, 1.44]\\
Percentage Change (RotRem vs No RotRem) & 41.5\% & {[}38.6\%, 44.4\%]\\
sd(Intercept) [molid:spin] & 0.66 & {[}0.62, 0.71]\\
\end{tabular}
\end{table}
\subsubsection{Combined model}
\label{sec:org9dc5f69}
To use all the data generated and draw the inferences in the main text, we fit a full hierarchical model.

\begin{verbatim}
 Family: negbinomial
  Links: mu = log; shape = identity
Formula: pes_calls ~ dimer_rot * rot_removal + (1 | mol_id:spin)
   Data: data (Number of observations: 1805)
  Draws: 4 chains, each with iter = 4000; warmup = 1000; thin = 1;
         total post-warmup draws = 12000

Multilevel Hyperparameters:
~mol_id:spin (Number of levels: 458)
              Estimate Est.Error l-95%
sd(Intercept)     0.63      0.02     0.59     0.67 1.01      314     1104

Regression Coefficients:
                              Estimate Est.Error l-95%
Intercept                         5.69      0.03     5.63     5.75 1.02      173      333
dimer_rotlbfgs                    0.03      0.01     0.01     0.04 1.00     7773     8783
rot_removalyes                    0.37      0.01     0.35     0.38 1.00     7986     8829
dimer_rotlbfgs:rot_removalyes     0.01      0.01    -0.01     0.04 1.00     6672     8277

Further Distributional Parameters:
      Estimate Est.Error l-95%
shape    63.99      2.75    58.70    69.59 1.00    11187     8962

Draws were sampled using sample(hmc). For each parameter, Bulk_ESS
and Tail_ESS are effective sample size measures, and Rhat is the potential
scale reduction factor on split chains (at convergence, Rhat = 1).
\end{verbatim}

The combined model reinforces the findings from the subset models and provides a more nuanced understanding of the factors influencing the number of PES calls.

The main effect for the L-BFGS optimizer (\texttt{dimer\_rotlbfgs}) is estimated at 0.03 (95\% CI: [0.01, 0.04]), consistent with the RotOptimizer subset model. This confirms that, holding rotation removal constant, the L-BFGS optimizer is associated with a small but discernible increase in \texttt{pes\_calls} compared to the Conjugate Gradient (CG) optimizer.

Similarly, the main effect for enabling the removal of external rotations (\texttt{rot\_removalyes}) is 0.37 (95\% CI: [0.35, 0.38]). This indicates a substantial and statistically significant increase in \texttt{pes\_calls} when rotation removal is active, compared to when it is not, holding the choice of optimizer constant. This represents a performance cost of approximately 45\%.

Crucially, the interaction term \texttt{dimer\_rotlbfgs:rot\_removalyes} is very small (0.01) and its 95\% CI ([-0.01, 0.04]) overlaps with zero. This suggests that there is no discernible interaction effect. In practical terms, the slight performance decrease from using L-BFGS instead of CG is consistent, regardless of whether external rotation removal is used or not.

The random intercept standard deviation remains large at 0.63 (95\% CI: [0.59, 0.67]), again highlighting significant baseline variability in \texttt{pes\_calls} across different molecular systems. The shape parameter is estimated at 63.99, indicating moderate overdispersion in the data.

\begin{figure}[htbp]
\centering
\includegraphics[scale=0.8]{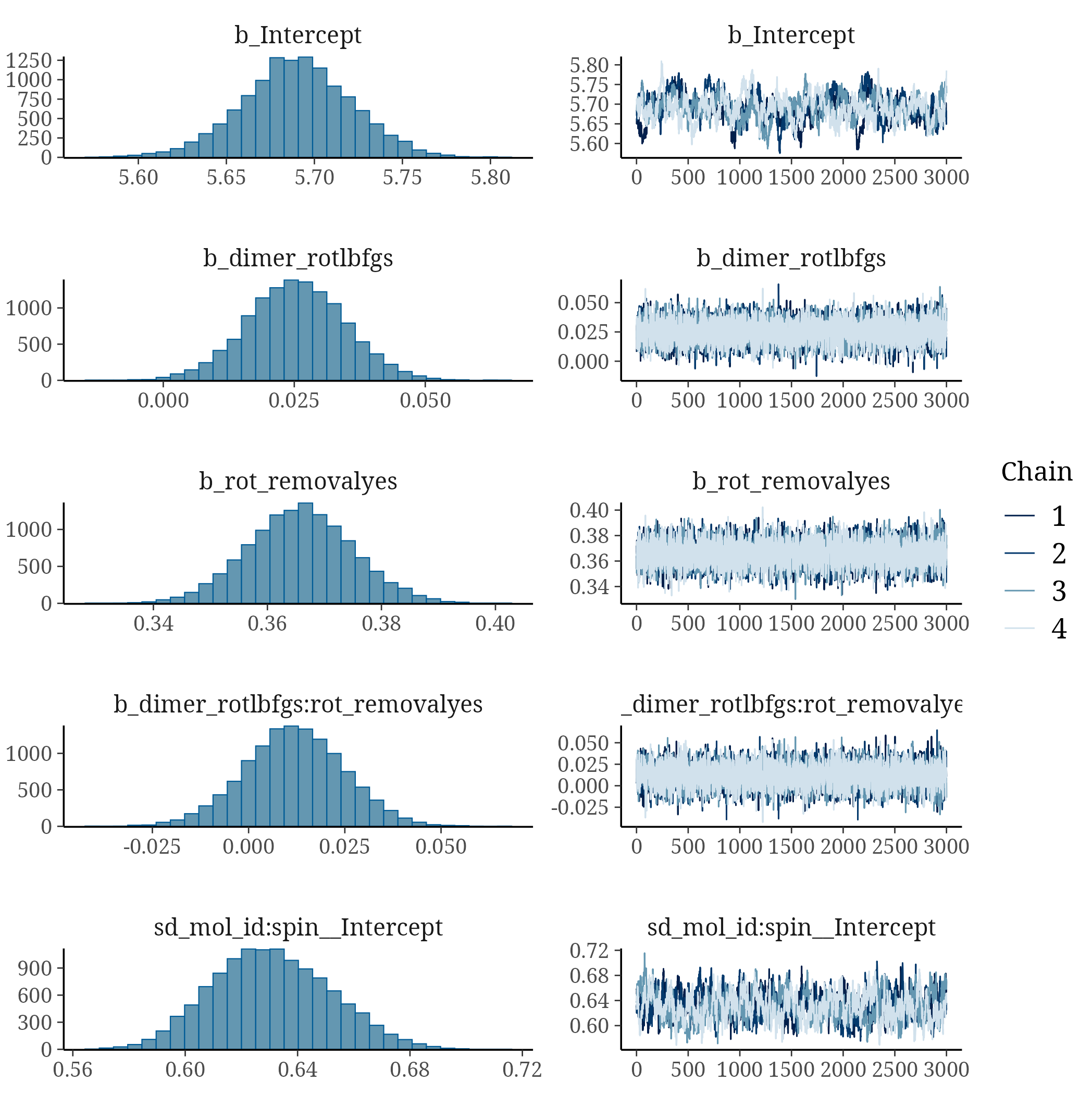}
\caption{\label{fig:brmsp_idall_traceplots}Trace plots for the main parameters of the combined performance model showing well-mixed chains.}
\end{figure}

Model diagnostics were excellent. All \texttt{Rhat} values are close to 1.0, and the \texttt{Bulk\_ESS} and \texttt{Tail\_ESS} values are large, indicating that the Markov chains converged and mixed well. Trace plots (Figure \ref{fig:brmsp_idall_traceplots}) confirmed good mixing of the chains.

\begin{figure}[htbp]
\centering
\includegraphics[scale=0.5]{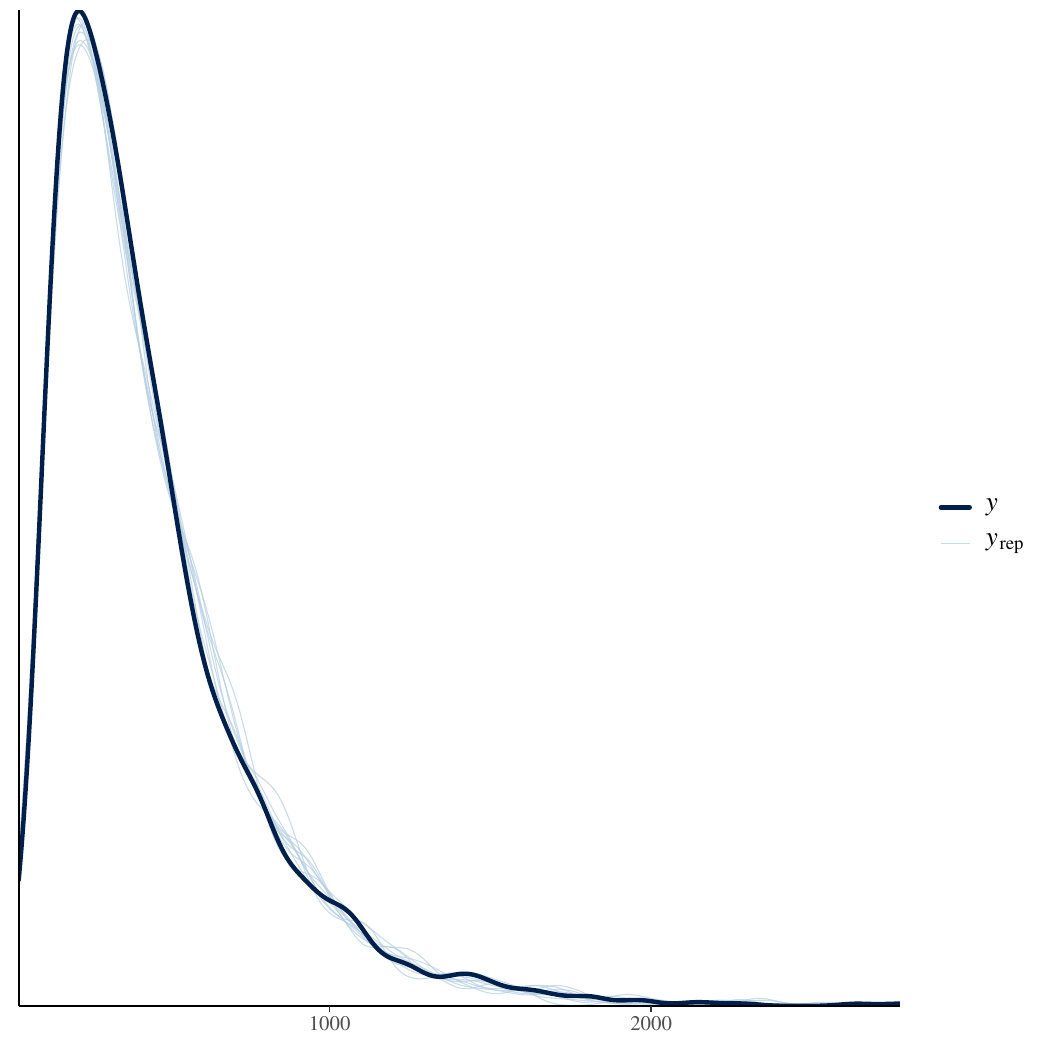}
\caption{\label{fig:brmsp_idall_pp}Posterior predictive check for the combined performance model (continuous).}
\end{figure}

Posterior predictive checks (Figure \ref{fig:brmsp_idall_pp}) indicate a good model fit, with the posterior predictive distribution closely matching the observed distribution of \texttt{pes\_calls}.

\begin{figure}[htbp]
\centering
\includegraphics[scale=0.5]{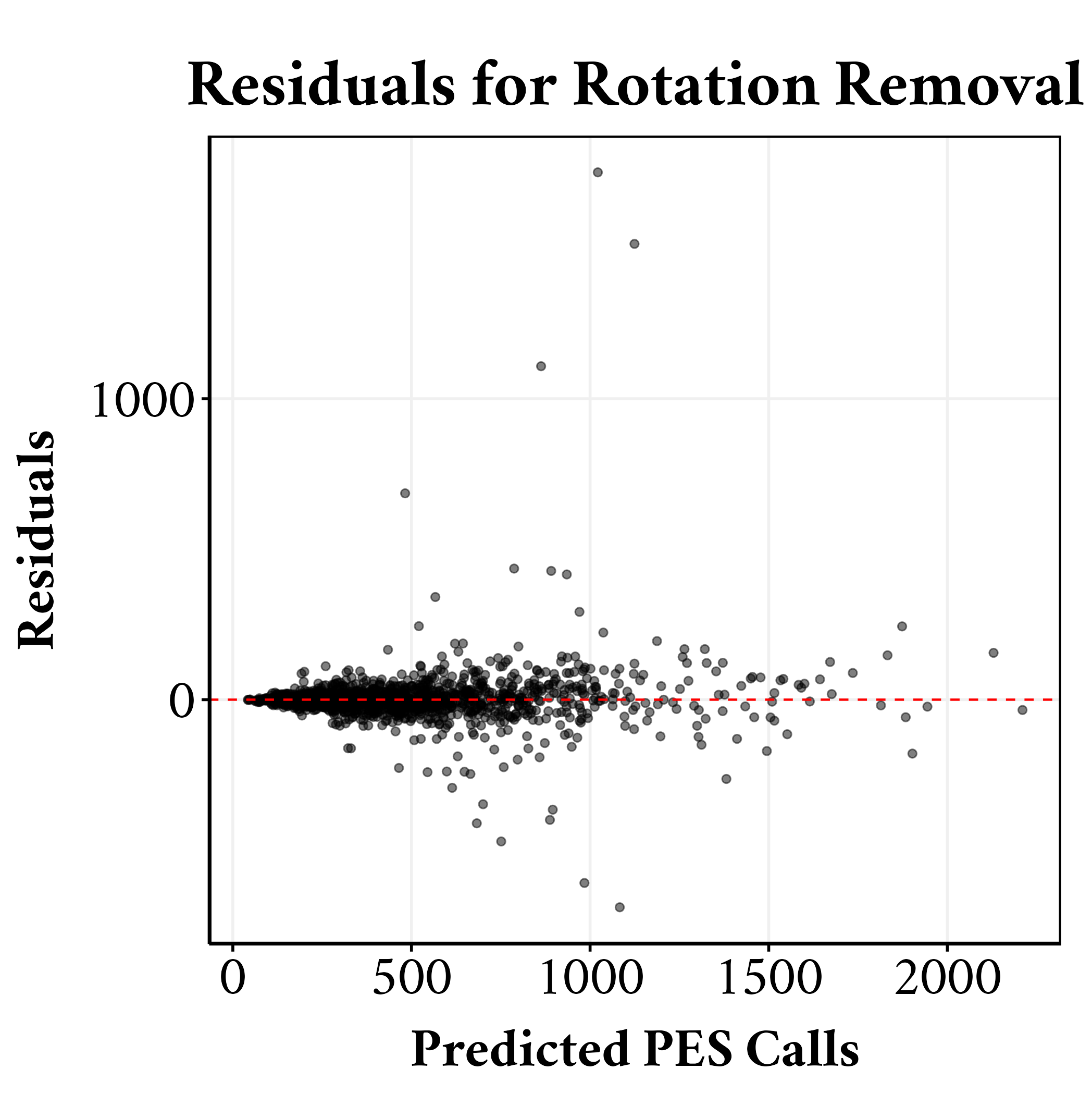}
\caption{\label{fig:brmsp_all_resid}Error residuals for the combined performance model.}
\end{figure}

The residual error analysis in Figure \ref{fig:brmsp_all_resid} shows slight heteroscedasticity, but less than the optimizer subset.

\begin{figure}[htbp]
\centering
\includegraphics[scale=0.5]{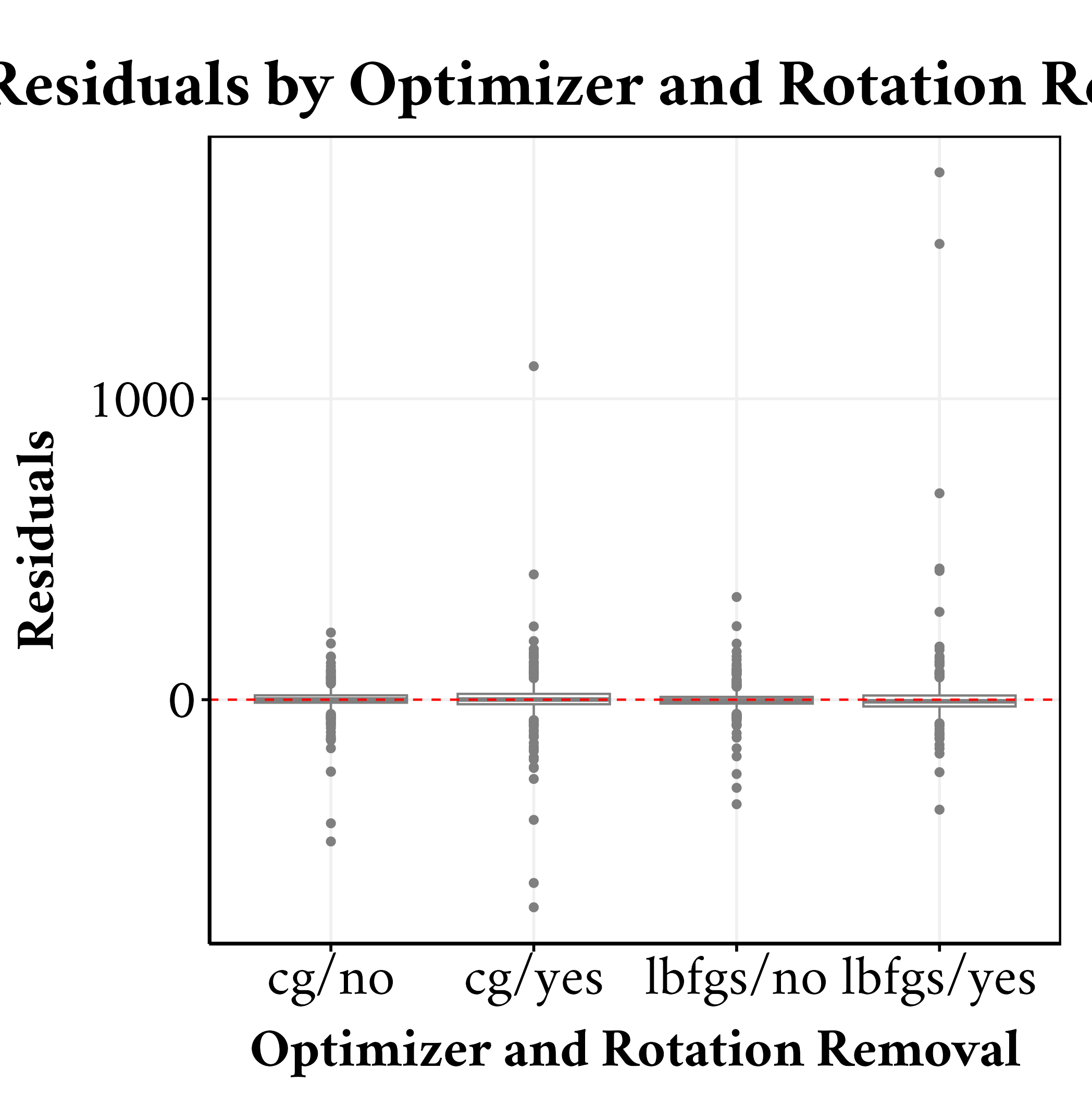}
\caption{\label{fig:brmsp_idall_boxplt}Boxplot of the error residuals for the combined performance model, showing no bias.}
\end{figure}

As before, the box plot in \ref{fig:brmsp_idall_boxplt} indicates that there is no systematic bias, so it is good enough for relative comparisons.

The posterior effects including the random effects are summarized in Table \ref{all_perf}.

\begin{table}[htbp]
\caption{\label{all_perf}Performance model effects, including credible intervals.}
\centering
\begin{tabular}{lrl}
Effect Type & Median Effect & 95\% CrI\\
\hline
Expected PES Calls (CG, No RotRem) & 295.90 & {[}278.0, 314.6]\\
Multiplicative Factor (L-BFGS vs CG / No RotRem) & 1.03 & {[}1.01, 1.04]\\
Percentage Change (L-BFGS vs CG / No RotRem) & 2.6\% & {[}0.7\%, 4.5\%]\\
Multiplicative Factor (RotRem Yes vs No / CG) & 1.44 & {[}1.42, 1.47]\\
Percentage Change (RotRem Yes vs No / CG) & 44.2\% & {[}41.6\%, 46.8\%]\\
Multiplicative Factor (Interaction) & 1.01 & {[}0.99, 1.04]\\
Percentage Change (Interaction) & 1.2\% & {[}-1.4\%, 3.8\%]\\
sd(Intercept) [molid:spin] & 0.63 & {[}0.59, 0.67]\\
\end{tabular}
\end{table}

In summary, the full model provides robust evidence that for optimizing the number of \texttt{pes\_calls}, the Conjugate Gradient (CG) optimizer should be preferred over L-BFGS, and external rotation removal should be disabled. The absence of an interaction effect simplifies the conclusion: these two choices are independent, and the optimal combination is CG without rotation removal.
\subsection{Total time model}
\label{sec:orgf9343e9}
Here we demonstrate the result of a combined total time model. As noted in the main text, the interpretation of this model shows that it is largely a proxy for the PES call performance, despite the difference in model family (Gamma compared to NegBinomial, due to continuous time data over discrete PES calls).

\begin{verbatim}
 Family: gamma
  Links: mu = log; shape = identity
Formula: tot_time ~ dimer_rot * rot_removal + (1 | mol_id:spin)
   Data: data (Number of observations: 1805)
  Draws: 4 chains, each with iter = 4000; warmup = 1000; thin = 1;
         total post-warmup draws = 12000

Multilevel Hyperparameters:
~mol_id:spin (Number of levels: 458)
              Estimate Est.Error l-95%
sd(Intercept)     0.66      0.02     0.61     0.70 1.01      326      687

Regression Coefficients:
                              Estimate Est.Error l-95%
Intercept                         3.10      0.03     3.03     3.16 1.03      169
dimer_rotlbfgs                    0.03      0.01     0.01     0.04 1.00     8956
rot_removalyes                    0.36      0.01     0.34     0.38 1.00     8554
dimer_rotlbfgs:rot_removalyes     0.01      0.01    -0.02     0.04 1.00     7661
                              Tail_ESS
Intercept                          288
dimer_rotlbfgs                    9686
rot_removalyes                    9371
dimer_rotlbfgs:rot_removalyes     8711

Further Distributional Parameters:
      Estimate Est.Error l-95%
shape    51.85      1.97    48.06    55.80 1.00    10773     9266

Draws were sampled using sample(hmc). For each parameter, Bulk_ESS
and Tail_ESS are effective sample size measures, and Rhat is the potential
scale reduction factor on split chains (at convergence, Rhat = 1).
\end{verbatim}

The analysis of the Gamma model for total wall time (\texttt{tot\_time}) reveals a narrative remarkably consistent with the pes\textsubscript{calls} performance model. The estimated coefficients for the predictors are nearly identical to those in the negative binomial model.

The effect for \texttt{dimer\_rotlbfgs} is 0.03 (95\% CI: [0.01, 0.04]), and for \texttt{rot\_removalyes} it is 0.36 (95\% CI: [0.34, 0.38]). As with the pes\textsubscript{calls} model, the interaction term remains small and not statistically discernible, with a coefficient of 0.01 and a 95\% CI of [-0.02, 0.04] that comfortably includes zero. This striking similarity confirms that the total wall time is predominantly driven by the number of PES calls required for convergence.

The random intercept standard deviation of 0.66 (95\% CI: [0.61, 0.70]) is also very similar to the combined \texttt{pes\_calls} model, indicating that molecules that are intrinsically more difficult (requiring more PES calls) also take proportionally longer to compute.

\begin{figure}[htbp]
\centering
\includegraphics[scale=0.8]{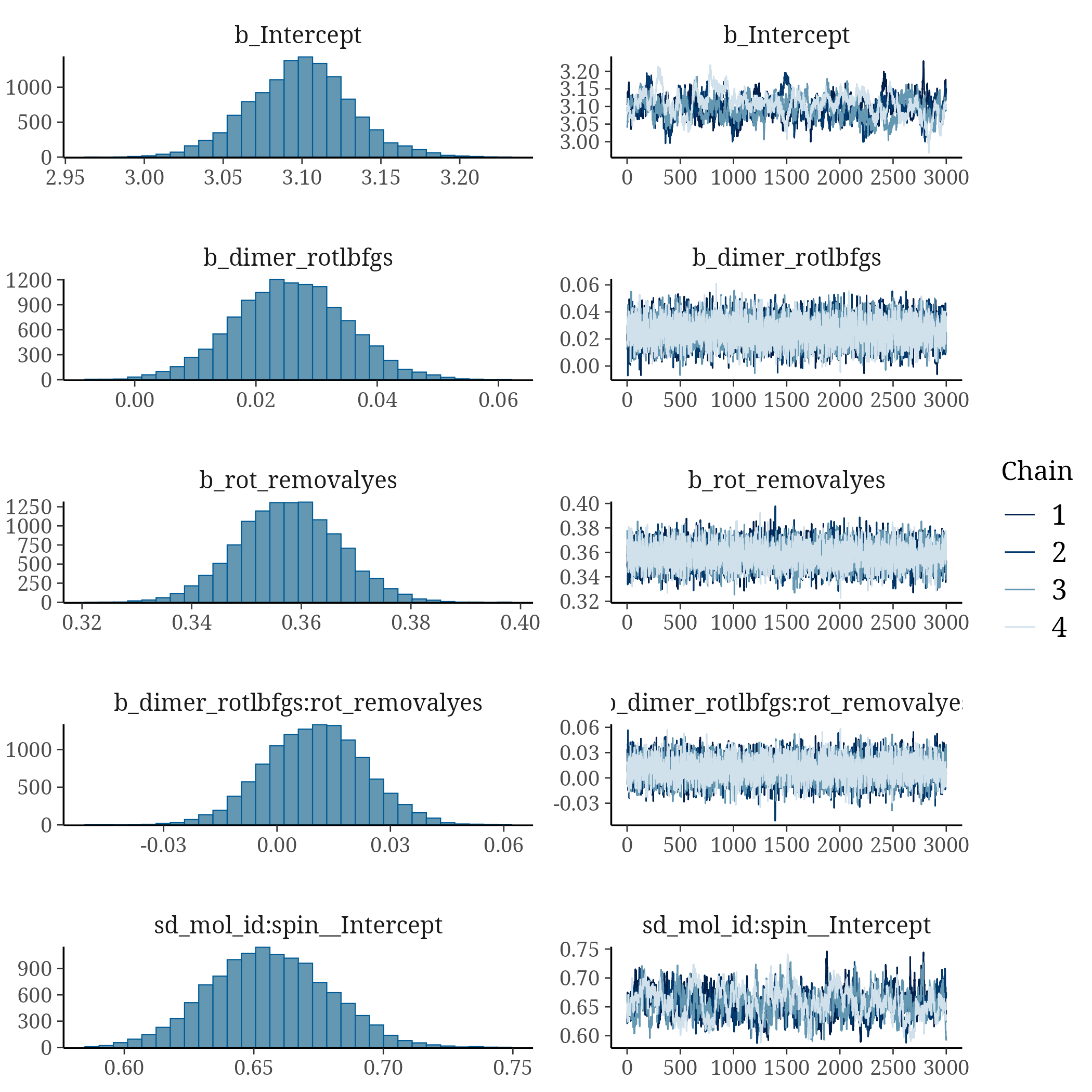}
\caption{\label{fig:brmst_idall_traceplots}Trace plots for the main parameters of the combined time model showing well-mixed chains.}
\end{figure}

Model diagnostics were excellent, as before. All \texttt{Rhat} values are close to 1.0, and the \texttt{Bulk\_ESS} and \texttt{Tail\_ESS} values are large, indicating that the Markov chains converged and mixed well. Trace plots (Figure \ref{fig:brmst_idall_traceplots}) confirmed good mixing of the chains.

\begin{figure}[htbp]
\centering
\includegraphics[scale=0.5]{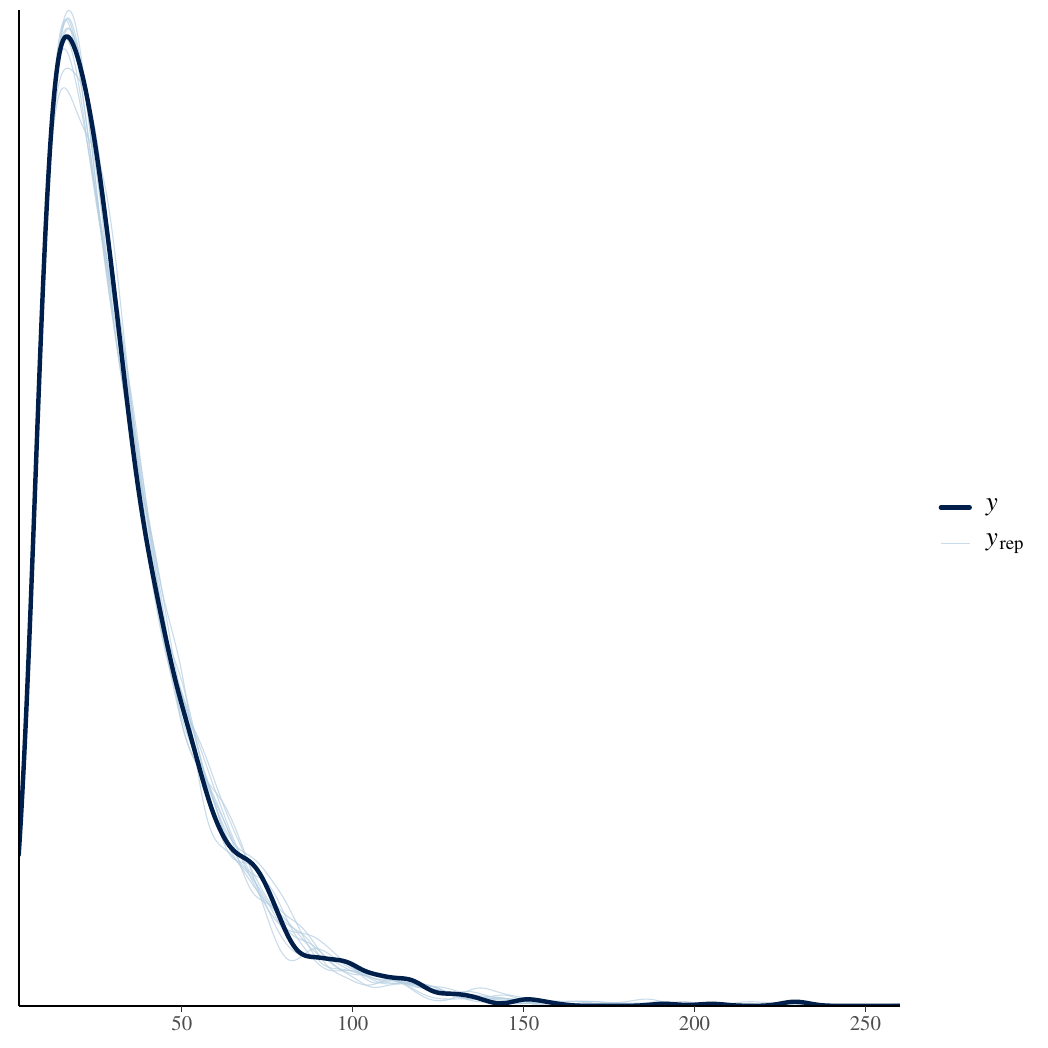}
\caption{\label{fig:brmst_idall_pp}Posterior predictive check for the combined time model (continuous).}
\end{figure}

Posterior predictive checks (Figure \ref{fig:brmst_idall_pp}) indicate a good model fit, with the posterior predictive distribution closely matching the observed distribution of \texttt{tot\_time}.

\begin{figure}[htbp]
\centering
\includegraphics[scale=0.5]{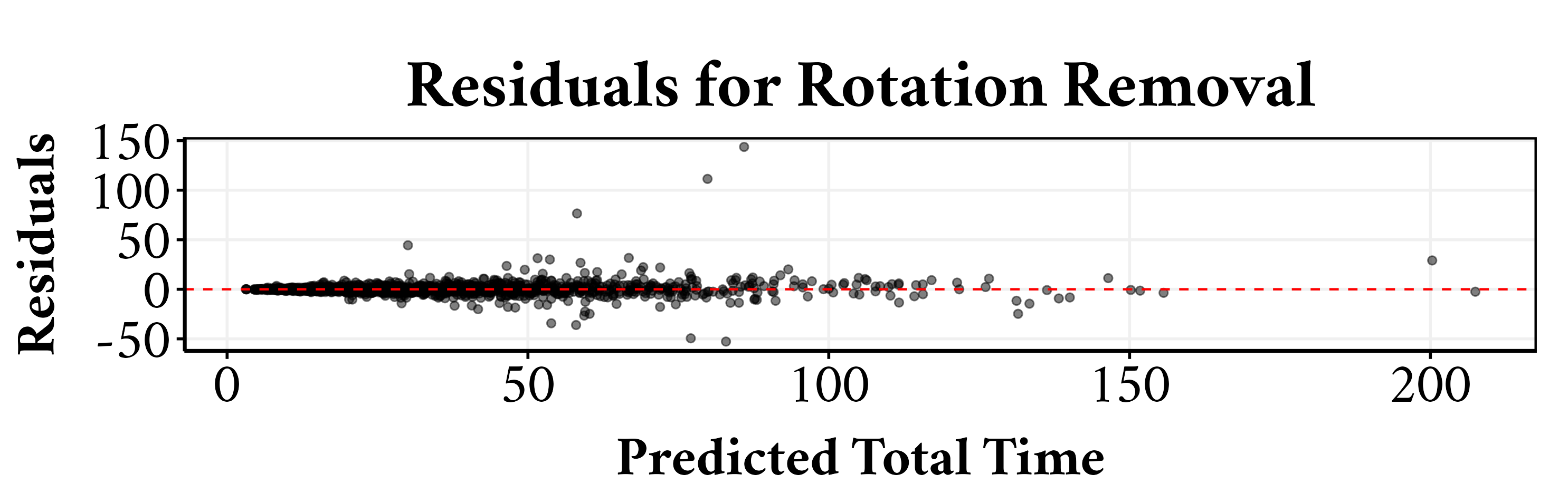}
\caption{\label{fig:brmst_all_resid}Error residuals for the combined time model.}
\end{figure}

The residual error analysis in Figure \ref{fig:brmst_all_resid} shows slight heteroscedasticity, but less than the optimizer subset.

\begin{figure}[htbp]
\centering
\includegraphics[scale=0.5]{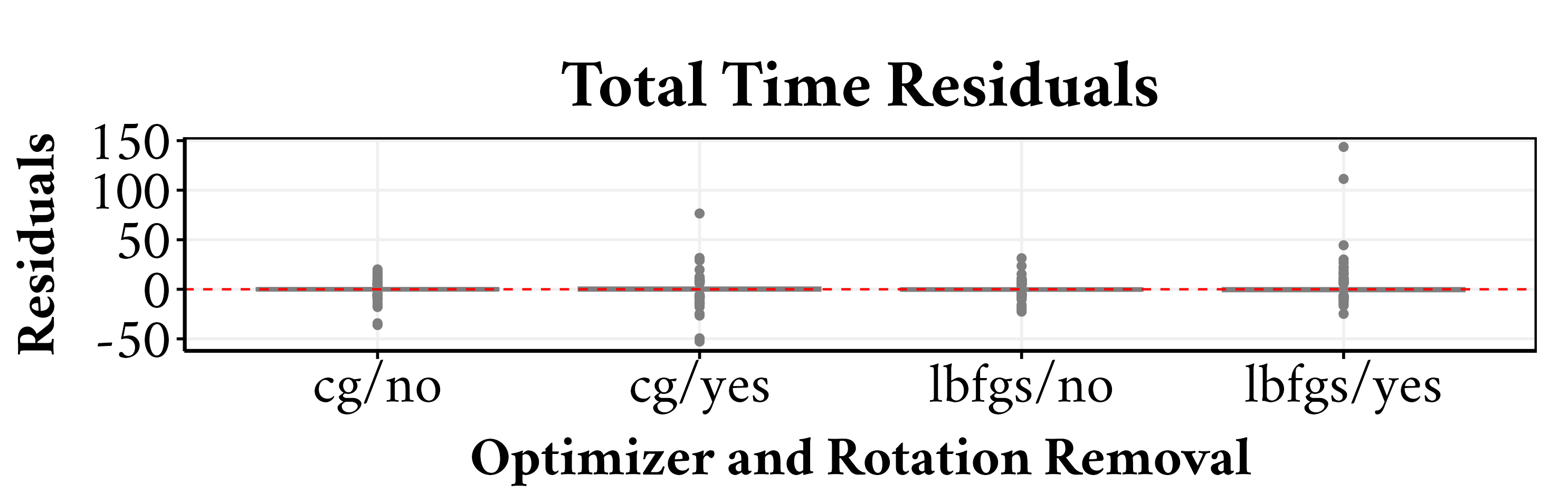}
\caption{\label{fig:brmst_idall_boxplt}Boxplot of the error residuals for the combined time model, showing no bias.}
\end{figure}

As before, the box plot in \ref{fig:brmst_idall_boxplt} indicates that there is no systematic bias, so it is good enough for relative comparisons.

The posterior effects including the random effects are summarized in Table \ref{all_ttime}.

\begin{table}[htbp]
\caption{\label{all_ttime}Total time model effects, including credible intervals.}
\centering
\begin{tabular}{lrl}
Effect Type & Median Effect & 95\% CrI\\
\hline
Expected Wall Time (CG, No RotRem) & 22.20 & {[}20.8, 23.7]\\
Multiplicative Factor (L-BFGS vs CG / No RotRem) & 1.03 & {[}1.01, 1.04]\\
Percentage Change (L-BFGS vs CG / No RotRem) & 2.6\% & {[}0.7\%, 4.5\%]\\
Multiplicative Factor (RotRem Yes vs No / CG) & 1.43 & {[}1.40, 1.46]\\
Percentage Change (RotRem Yes vs No / CG) & 43.0\% & {[}40.4\%, 45.6\%]\\
Multiplicative Factor (Interaction) & 1.01 & {[}0.98, 1.04]\\
Percentage Change (Interaction) & 1.0\% & {[}-1.6\%, 3.6\%]\\
sd(Intercept) [molid:spin] & 0.66 & {[}0.61, 0.70]\\
\end{tabular}
\end{table}

This model serves as a real-world validation of the conclusions drawn from the performance model. The most efficient method in terms of computational steps (CG optimizer, no rotation removal) is also the fastest in terms of total wall time.
\subsection{Success modeling}
\label{sec:org5b36fda}
To assess the relative reliability of each setting, we model the probability of a successful calculation using a hierarchical logistic regression model. Success is a binary outcome (1 for success, 0 for failure).
\subsubsection{Optimizer for the rotation phase}
\label{sec:orgd69812f}
Here we use the (RotOptimizer) subset, i.e. with no rotation removal.

\begin{verbatim}
 Family: bernoulli
  Links: mu = logit
Formula: success ~ dimer_rot + (1 | mol_id:spin)
   Data: data (Number of observations: 1000)
  Draws: 4 chains, each with iter = 4000; warmup = 1000; thin = 1;
         total post-warmup draws = 12000

Multilevel Hyperparameters:
~mol_id:spin (Number of levels: 500)
              Estimate Est.Error l-95%
sd(Intercept)     2.61      0.55     1.67     3.81 1.00     2719     5242

Regression Coefficients:
               Estimate Est.Error l-95%
Intercept          5.80      0.84     4.40     7.72 1.00     3336     5181
dimer_rotlbfgs    -1.25      0.39    -2.06    -0.52 1.00    11747     8823

Draws were sampled using sample(hmc). For each parameter, Bulk_ESS
and Tail_ESS are effective sample size measures, and Rhat is the potential
scale reduction factor on split chains (at convergence, Rhat = 1).
\end{verbatim}

In absolute terms, both optimizers are highly reliable. Based on the model's median intercept, the typical probability of success for the baseline CG optimizer is extremely high at \(\approx\) 99.7\%. For the L-BFGS optimizer, the success rate also remains very high at \(\approx\) 99.0\%.

Despite these high absolute rates, the statistical model is detects a robust relative difference between the two. The coefficient for \texttt{dimer\_rotlbfgs} is -1.25 (95\% CI: [-2.06, -0.52]), indicating a statistically discernible difference in the log-odds of success.

\begin{figure}[htbp]
\centering
\includegraphics[scale=0.8]{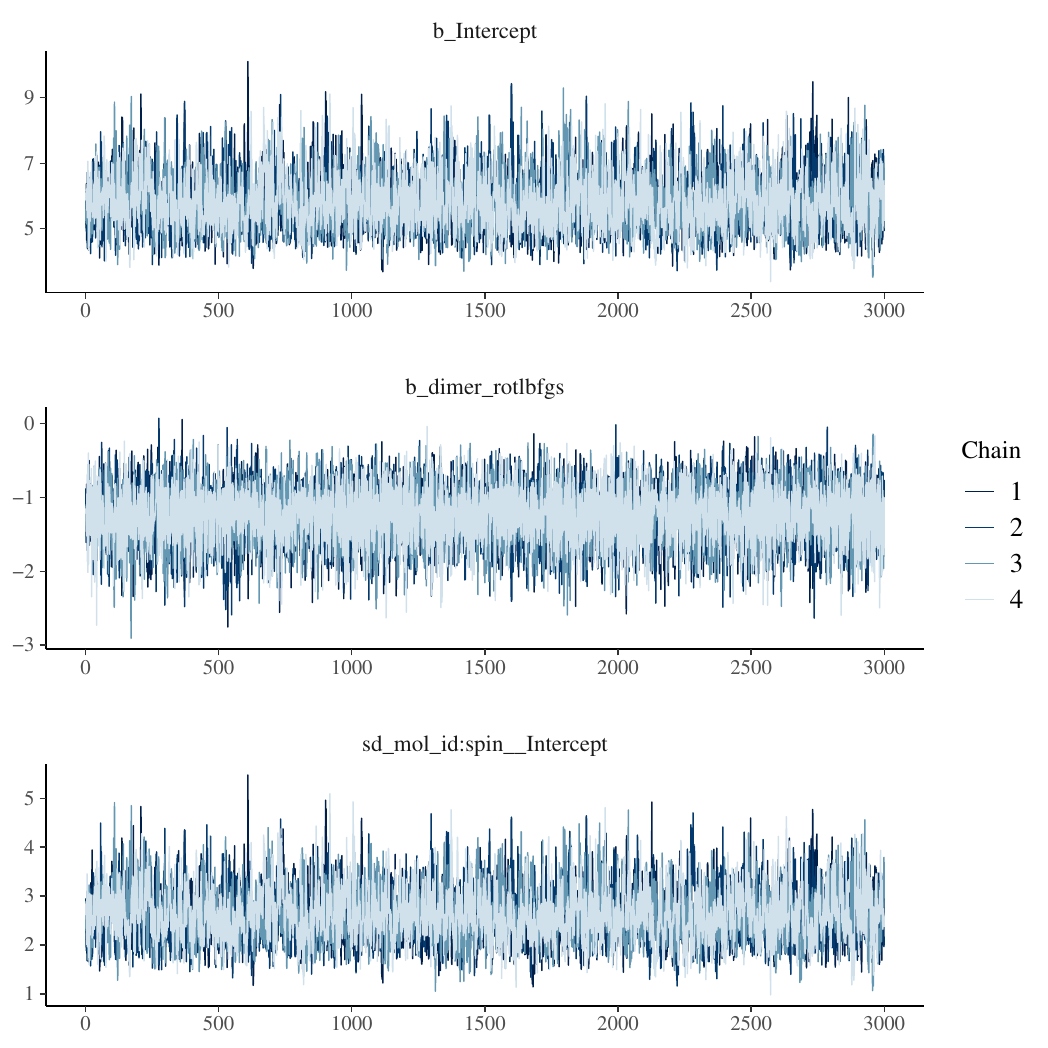}
\caption{\label{fig:brmss_rotopt_traceplots}Trace plots for the main parameters of the RotOptimizer success model, showing well-mixed chains.}
\end{figure}

Model diagnostics were excellent. All \texttt{Rhat} values are close to 1.0, and the \texttt{Bulk\_ESS} and \texttt{Tail\_ESS} values are large, indicating that the Markov chains converged and mixed well. Trace plots (Figure \ref{fig:brmss_rotopt_traceplots}) confirmed good mixing of the chains.

\begin{figure}[htbp]
\centering
\includegraphics[scale=0.5]{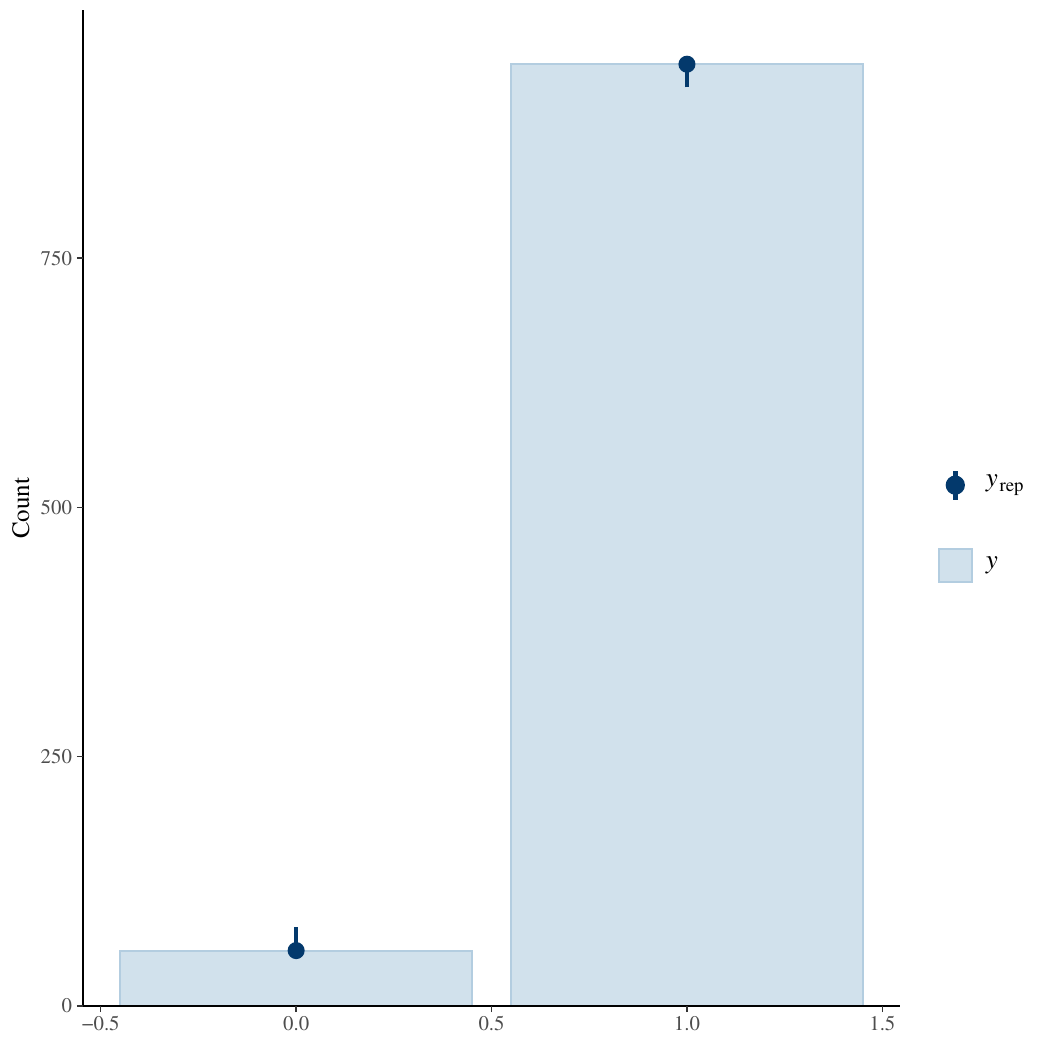}
\caption{\label{fig:brmss_rotopt_pp}Posterior predictive check for the RotOptimizer success model.}
\end{figure}

Posterior predictive checks (Figure \ref{fig:brmss_rotopt_pp}) indicate a good model fit, with the posterior predictive distribution closely matching the observed distribution of \texttt{success}.

\begin{figure}[htbp]
\centering
\includegraphics[scale=0.5]{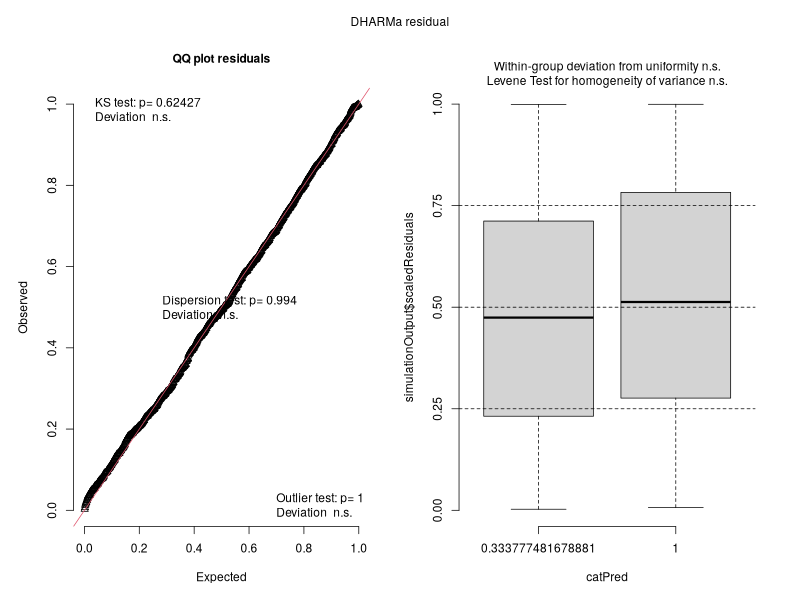}
\caption{\label{fig:brmss_rotopt_dharma}DHARMa diagnostic plots for the RotOptimizer success model, showing a good fit with no significant deviations in residuals.}
\end{figure}

Further diagnostics were performed using the DHARMa package, which creates standardized residuals for complex models. The results in Figure \ref{fig:brmss_rotopt_dharma} are excellent and support the validity of the model.

\begin{description}
\item[{QQ plot (left panel)}] This plot compares the empirical distribution of residuals against the expected uniform distribution. The points lie almost perfectly along the diagonal line, indicating that the model's residuals are distributed as expected. This visual finding is supported by formal statistical tests for correct distribution (Kolmogorov-Smirnov test, p=0.62), correct dispersion (p=0.99), and outliers (p=1), none of which show a significant deviation.
\item[{Residuals vs. Predictors (right panel)}] This plot checks for non-uniform patterns in the residuals across the two optimizer groups. The boxplots are centered around the expected median of 0.5 and have similar spreads. The non-significant Levene Test confirms that the variance of the residuals is homogeneous across the groups.
\end{description}

The practical meaning of this relative difference is best understood through the odds ratios summarized in Table \ref{rotopt_success}. While both methods nearly always succeed, the odds of success are substantially different. The odds ratio for L-BFGS relative to CG is just 0.29, which corresponds to a median change in the odds of success of -70.9\% (95\% CrI: [-87.2\%, -40.4\%]).

\begin{table}[htbp]
\caption{\label{rotopt_success}RotOptimizer success model effects, including credible intervals.}
\centering
\begin{tabular}{lrl}
Effect Type & Median Effect & 95\% CrI\\
\hline
Probability (CG Success) & 1.00 & {[}0.99, 1.00]\\
Odds Ratio (L-BFGS vs. CG) & 0.29 & {[}0.13, 0.60]\\
sd(Intercept) [molid:spin] & 2.56 & {[}1.67, 3.81]\\
\end{tabular}
\end{table}

Framed differently, an odds ratio of 0.29 indicates that the odds of failure, though small for both, are over three times higher for L-BFGS than for CG (1 / 0.29 \(\approx\) 3.45). Therefore, while both methods are very effective, the analysis shows that the CG optimizer is the more robust and reliable choice when external rotation is not being removed.
\subsubsection{Removal of external rotation}
\label{sec:org69555b8}
This model is fit to the RotRemoval subset, with the CG optimizer, as it is the better of the two from the previous section.

\begin{verbatim}
 Family: bernoulli
  Links: mu = logit
Formula: success ~ rot_removal + (1 | mol_id:spin)
   Data: data (Number of observations: 1000)
  Draws: 4 chains, each with iter = 4000; warmup = 1000; thin = 1;
         total post-warmup draws = 12000

Multilevel Hyperparameters:
~mol_id:spin (Number of levels: 500)
              Estimate Est.Error l-95%
sd(Intercept)     4.36      1.12     2.63     6.91 1.00     2997     4643

Regression Coefficients:
               Estimate Est.Error l-95%
Intercept          8.66      1.90     5.81    13.12 1.00     3157     4678
rot_removalyes     0.72      0.55    -0.31     1.85 1.00    14155     8596

Draws were sampled using sample(hmc). For each parameter, Bulk_ESS
and Tail_ESS are effective sample size measures, and Rhat is the potential
scale reduction factor on split chains (at convergence, Rhat = 1).
\end{verbatim}

The key finding from this model is that there is no evidence of a discernible effect. The coefficient for \texttt{rot\_removalyes} is 0.72, but its 95\% CI [-0.31, 1.85] broadly overlaps with zero. This indicates that we cannot confidently conclude that enabling rotation removal has any impact—positive or negative—on the success rate.

It is also important to note that the baseline success rate (with the CG optimizer and no rotation removal) is already effectively 100\%, as indicated by the massive intercept of 8.66. When a process is already so reliable, it is difficult for a modification to show a measurable improvement.

Model diagnostics were excellent. All \texttt{Rhat} values are close to 1.0, and \texttt{Bulk\_ESS} and \texttt{Tail\_ESS} values are large, indicating good chain convergence.

\begin{figure}[htbp]
\centering
\includegraphics[scale=0.8]{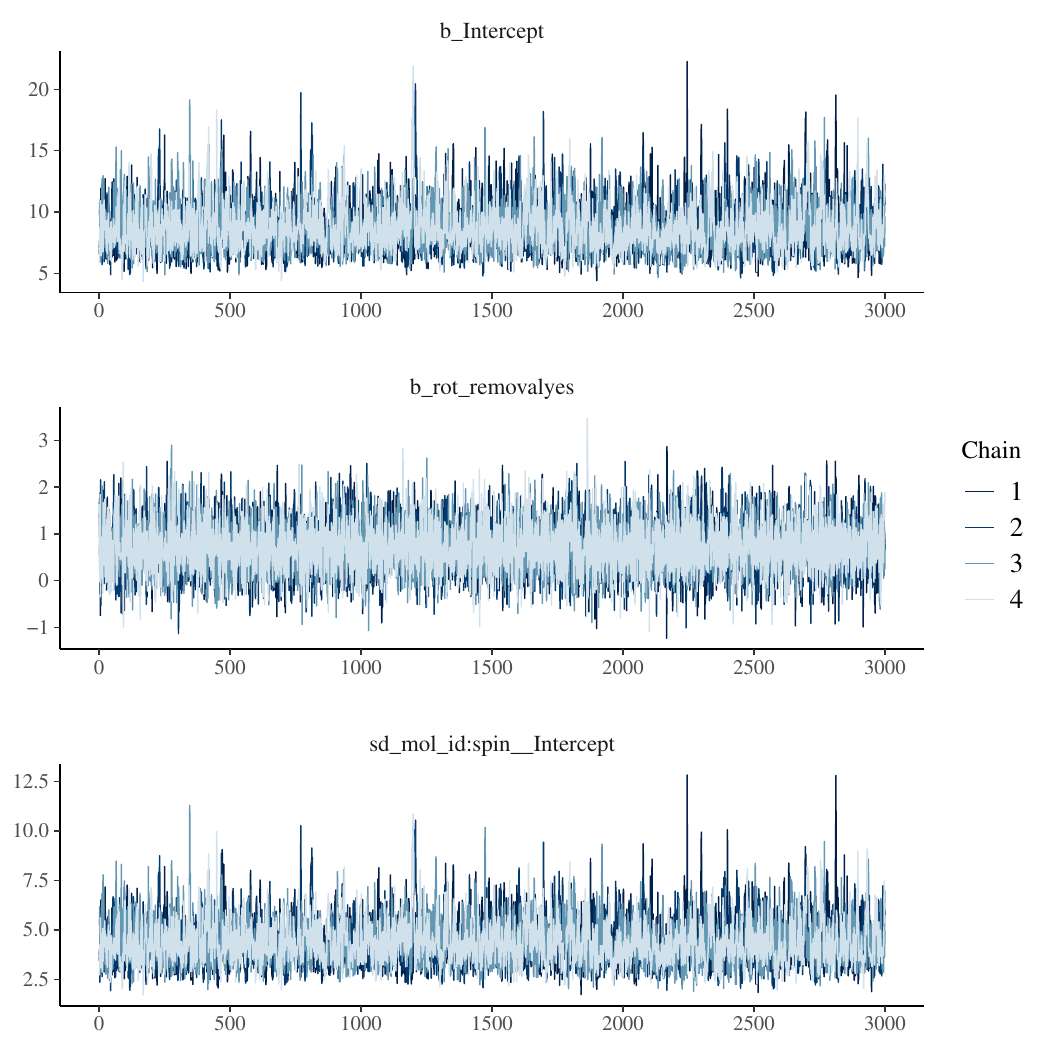}
\caption{\label{fig:brmss_rotrem_traceplots}Trace plots for the main parameters of the RotRemoval success model, showing well-mixed chains.}
\end{figure}

\begin{figure}[htbp]
\centering
\includegraphics[scale=0.5]{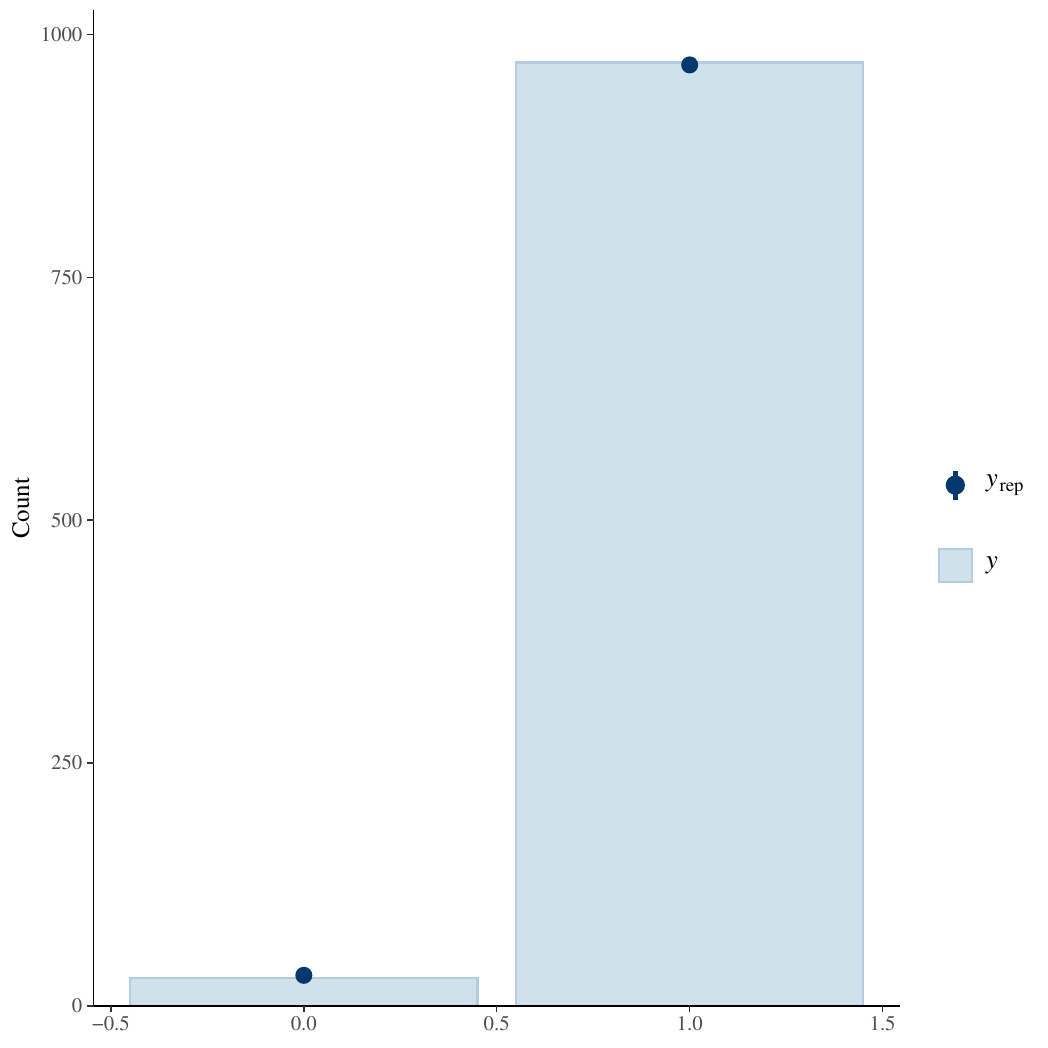}
\caption{\label{fig:brmss_rotrem_pp}Posterior predictive check for the RotRemoval success model.}
\end{figure}

Posterior predictive checks (Figure \ref{fig:brmss_rotrem_pp}) indicate a good model fit, with the posterior predictive distribution closely matching the observed distribution of \texttt{success}.

\begin{figure}[htbp]
\centering
\includegraphics[scale=0.5]{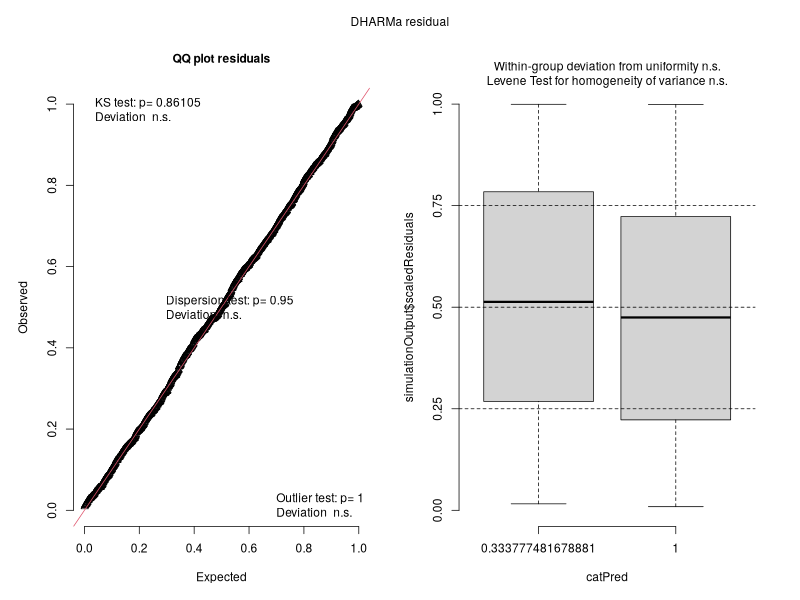}
\caption{\label{fig:brmss_rotrem_dharma}DHARMa diagnostic plots for the RotRemoval success model, showing a good fit.}
\end{figure}

The DHARMa diagnostic plots in Figure \ref{fig:brmss_rotrem_dharma} further support the model's validity. The QQ plot's points fall along the diagonal, and the formal tests for deviation are all non-significant (e.g., KS test, p=0.86; Dispersion test, p=0.95), indicating correctly distributed residuals.

This statistical uncertainty is clearly reflected in the summary table (Table \ref{rotrem_success}). While the median Odds Ratio for using rotation removal is 2.0, suggesting it might double the odds of success, the 95\% credible interval [0.73, 6.35] is extremely wide. Since the interval contains 1.0 (no change in odds), we cannot rule out the possibility of no effect or even a slight negative effect.

\begin{table}[htbp]
\caption{\label{rotrem_success}RotRemoval success model effects, including credible intervals.}
\centering
\begin{tabular}{lrl}
Effect Type & Median Effect & 95\% CrI\\
\hline
Probability (CG Success) & 1.0 & {[}1.00, 1.00]\\
Odds Ratio RotRemove (Yes vs. No) & 2.0 & {[}0.73, 6.35]\\
sd(Intercept) [molid:spin] & 3.6 & {[}2.79, 4.80]\\
\end{tabular}
\end{table}

In sharp contrast to the clear and practical impact of the optimizer choice, this analysis provides no evidence to suggest that enabling or disabling external rotation removal has a discernible effect on the ultimate success of a calculation.
\subsubsection{Combined model}
\label{sec:org7374f1d}
To verify the findings from the subset models and test for any interaction effects, a full model incorporating both the optimizer choice and rotation removal setting was fit to all available data.

\begin{verbatim}
 Family: bernoulli
  Links: mu = logit
Formula: success ~ dimer_rot * rot_removal + (1 | mol_id:spin)
   Data: data (Number of observations: 2000)
  Draws: 4 chains, each with iter = 4000; warmup = 1000; thin = 1;
         total post-warmup draws = 12000

Multilevel Hyperparameters:
~mol_id:spin (Number of levels: 500)
              Estimate Est.Error l-95%
sd(Intercept)     3.67      0.52     2.79     4.80 1.00     2970     5321

Regression Coefficients:
                              Estimate Est.Error l-95%
Intercept                         7.45      0.85     5.99     9.31 1.00     4031
dimer_rotlbfgs                   -1.60      0.42    -2.45    -0.81 1.00    10344
rot_removalyes                    0.65      0.49    -0.30     1.62 1.00    10602
dimer_rotlbfgs:rot_removalyes     0.06      0.60    -1.10     1.23 1.00     9521
                              Tail_ESS
Intercept                         5602
dimer_rotlbfgs                    9006
rot_removalyes                    9400
dimer_rotlbfgs:rot_removalyes     8970

Draws were sampled using sample(hmc). For each parameter, Bulk_ESS
and Tail_ESS are effective sample size measures, and Rhat is the potential
scale reduction factor on split chains (at convergence, Rhat = 1).
\end{verbatim}

The full model reinforces the conclusions drawn from the simpler subset models. The main effect for the optimizer choice, \texttt{dimer\_rotlbfgs}, is large, negative, and statistically discernible, with a coefficient of -1.60 (95\% CI: [-2.45, -0.81]). In contrast, the main effect for \texttt{rot\_removalyes} and the interaction term, \texttt{dimer\_rotlbfgs:rot\_removalyes}, are both small and have credible intervals that broadly overlap with zero.

This indicates that while the choice of optimizer has a significant impact on the reliability of the calculation, the decision to use rotation removal does not. Furthermore, the lack of a discernible interaction effect means that the performance penalty of using L-BFGS is consistent (though slightly higher reliability with rotation removal is indicated), regardless of the rotation removal setting. As before, all tested configurations show a very high absolute success rate ( >99\% ).

\begin{figure}[htbp]
\centering
\includegraphics[scale=0.8]{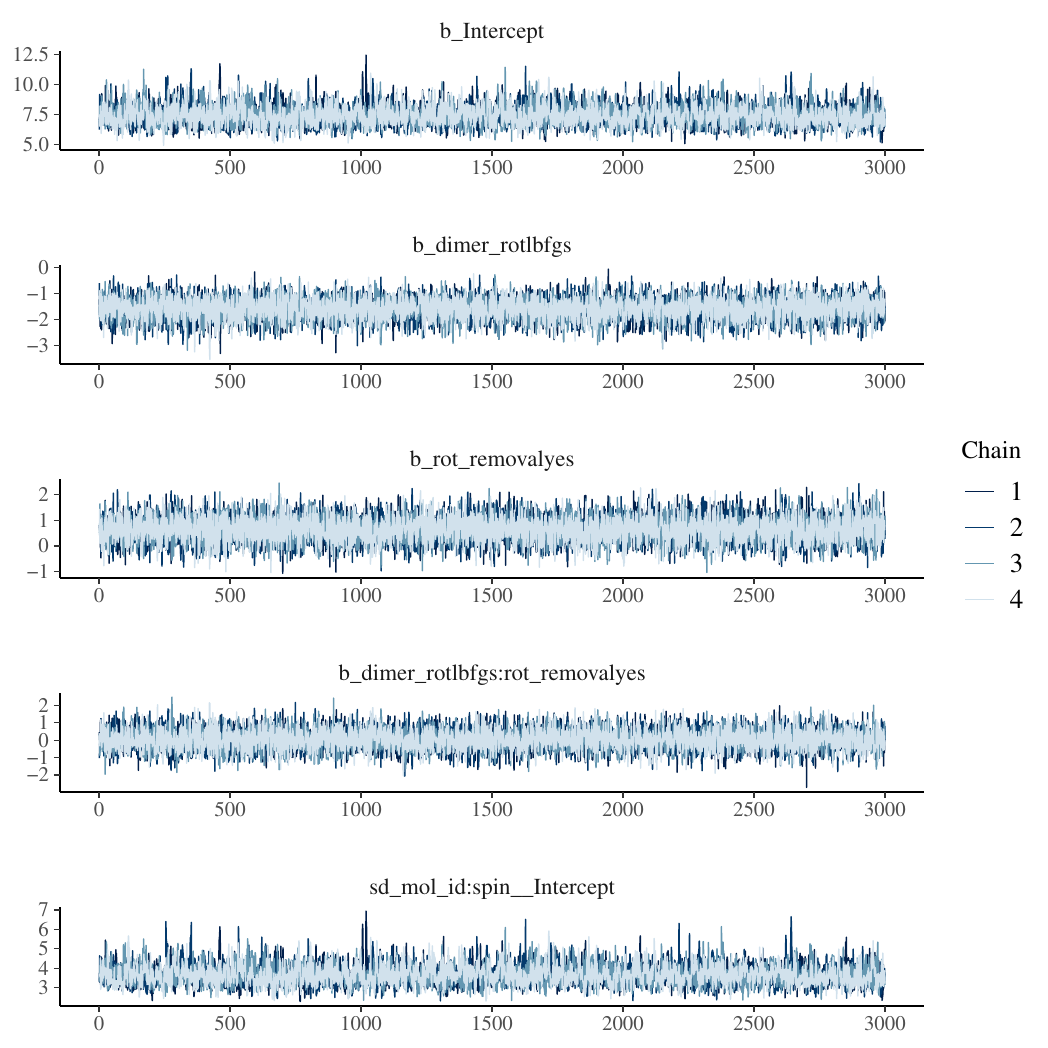}
\caption{\label{fig:brmss_all_traceplots}Trace plots for the main parameters of the combined success model, showing well-mixed chains.}
\end{figure}

Model diagnostics were excellent. All Rhat values are close to 1.0, and Bulk\textsubscript{ESS} and Tail\textsubscript{ESS} values are large, indicating good chain convergence. Trace plots (Figure \ref{fig:brmss_all_traceplots}) confirmed good mixing of the chains.

\begin{figure}[htbp]
\centering
\includegraphics[scale=0.5]{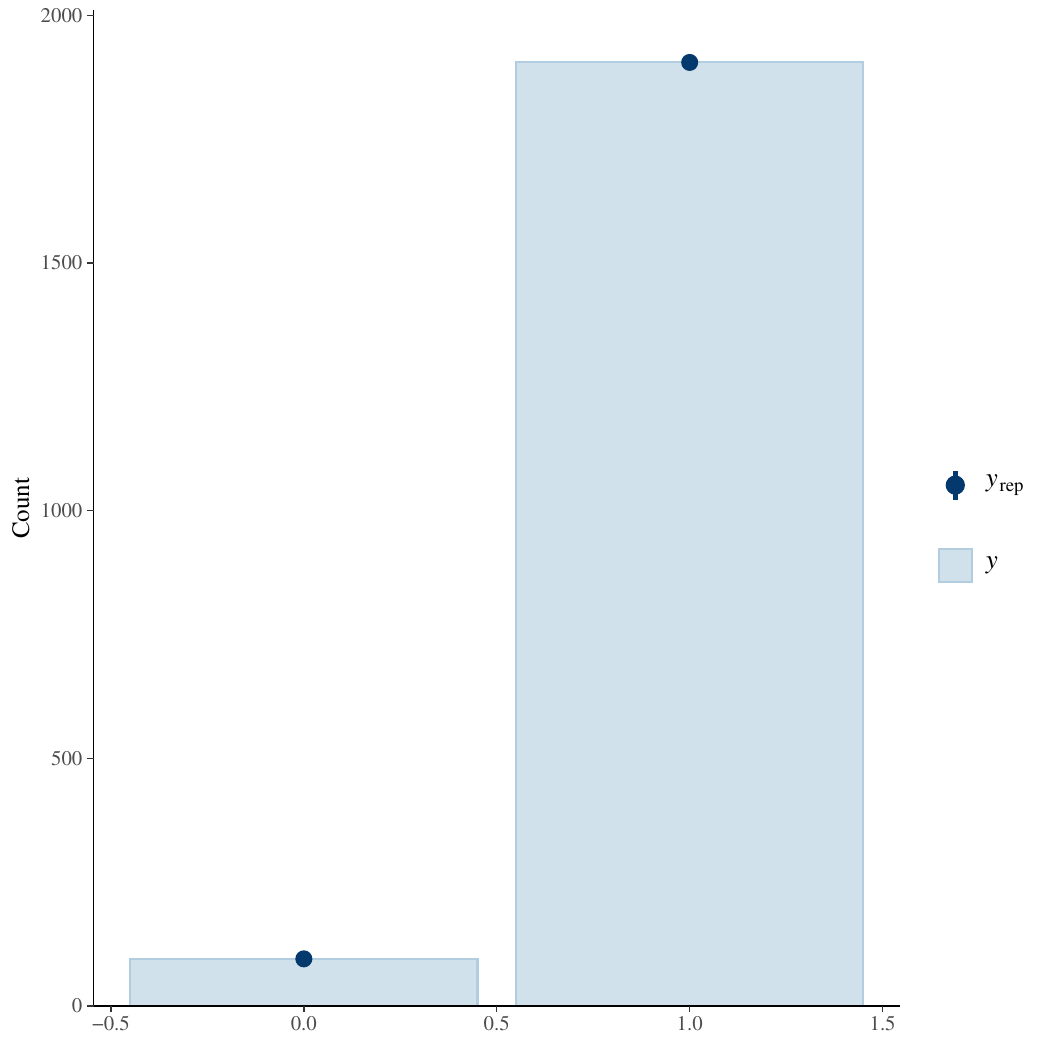}
\caption{\label{fig:brmss_all_pp}Posterior predictive check for the combined success model.}
\end{figure}

Posterior predictive checks (Figure \ref{fig:brmss_all_pp}) indicate a good model fit, with the posterior predictive distribution closely matching the observed distribution of success.

\begin{figure}[htbp]
\centering
\includegraphics[scale=0.5]{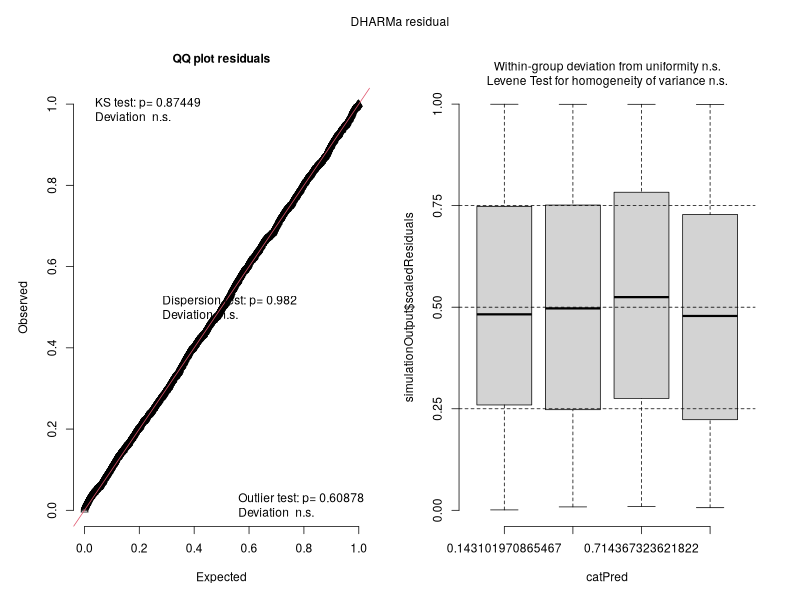}
\caption{\label{fig:brmss_all_dharma}DHARMa diagnostic plots for the combined success model, showing a good fit with no significant deviations in residuals.}
\end{figure}

Further diagnostics using the DHARMa package (Figure \ref{fig:brmss_all_dharma}) also support the validity of the model, with residuals behaving as expected and showing no significant patterns or deviations from uniformity.

The practical effects are summarized in Table \ref{all_success}. The Odds Ratio for using L-BFGS versus CG is 0.20, a substantial and discernible reduction in the odds of success. The odds ratios for rotation removal and the interaction term both have wide credible intervals that contain 1.0, confirming they are not statistically discernible.

A subtle question is whether rotation removal has a benefit specifically for the L-BFGS optimizer. The model's median estimate for this conditional effect (the sum of the rot\textsubscript{removalyes} and interaction effects) suggests that enabling rotation removal doubles the odds of success for L-BFGS. However, due to the high uncertainty in both of these parameters, this combined effect is also not statistically discernible and we cannot confidently claim it is a reliable improvement.

\begin{table}[htbp]
\caption{\label{all_success}Combined success model effects, including credible intervals.}
\centering
\begin{tabular}{lrl}
Effect Type & Median Effect & 95\% CrI\\
\hline
Probability (CG Success) & 1.0 & {[}1.00, 1.00]\\
Odds Ratio (L-BFGS vs. CG) & 0.2 & {[}0.09, 0.45]\\
Interaction Odds Ratio & 1.1 & {[}0.33, 3.42]\\
Odds Ratio RotRemove (Yes vs. No) & 1.9 & {[}0.74, 5.07]\\
sd(Intercept) [molid:spin] & 3.6 & {[}2.79, 4.80]\\
\end{tabular}
\end{table}

The combined success model provides robust, comprehensive evidence for a clear conclusion. For maximum reliability, the Conjugate Gradient (CG) optimizer should be used. The choice of enabling or disabling external rotation removal has no statistically discernible impact on the success rate. The optimal and most robust method is therefore CG, with or without rotation removal.

\end{appendices}
\bibliography{cg_lbfgs_stat,manual}
\end{document}

%% file: authors_arxiv.tex
\author{ \href{https://orcid.org/0000-0002-2393-8056}{\includegraphics[scale=0.06]{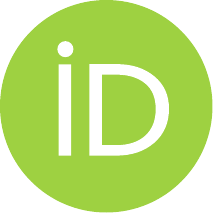}\hspace{1mm}Rohit Goswami}\thanks{Corresponding Author} \\
    Science Institute and Faculty of Physical Sciences \\
    University of Iceland, 107 Reykjavík, Iceland \\[1ex]
    \textit{Also at:} Department of Mechanical and Materials Engineering \\
    Queen’s University, 60 Union Street, Kingston, Ontario, Canada \\[1ex]
	\texttt{rgoswami@ieee.org} 
}

%% file: arxiv.bbl
\begin{thebibliography}{10}

\bibitem{apraNWChemPresentFuture2020}
{\sc Apr{\`a}, E., Bylaska, E.~J., De~Jong, W.~A., Govind, N., Kowalski, K.,
  Straatsma, T.~P., Valiev, M., Van~Dam, H. J.~J., Alexeev, Y., Anchell, J.,
  Anisimov, V., Aquino, F.~W., {Atta-Fynn}, R., Autschbach, J., Bauman, N.~P.,
  Becca, J.~C., Bernholdt, D.~E., {Bhaskaran-Nair}, K., Bogatko, S., Borowski,
  P., Boschen, J., Brabec, J., Bruner, A., Cau{\"e}t, E., Chen, Y., Chuev,
  G.~N., Cramer, C.~J., Daily, J., Deegan, M. J.~O., Dunning, T.~H., Dupuis,
  M., Dyall, K.~G., Fann, G.~I., Fischer, S.~A., Fonari, A., Fr{\"u}chtl, H.,
  Gagliardi, L., Garza, J., Gawande, N., Ghosh, S., Glaesemann, K., G{\"o}tz,
  A.~W., Hammond, J., Helms, V., Hermes, E.~D., Hirao, K., Hirata, S.,
  Jacquelin, M., Jensen, L., Johnson, B.~G., J{\'o}nsson, H., Kendall, R.~A.,
  Klemm, M., Kobayashi, R., Konkov, V., Krishnamoorthy, S., Krishnan, M., Lin,
  Z., Lins, R.~D., Littlefield, R.~J., Logsdail, A.~J., Lopata, K., Ma, W.,
  Marenich, A.~V., Martin Del~Campo, J., {Mejia-Rodriguez}, D., Moore, J.~E.,
  Mullin, J.~M., Nakajima, T., Nascimento, D.~R., Nichols, J.~A., Nichols,
  P.~J., Nieplocha, J., {Otero-de-la-Roza}, A., Palmer, B., Panyala, A.,
  Pirojsirikul, T., Peng, B., Peverati, R., Pittner, J., Pollack, L., Richard,
  R.~M., Sadayappan, P., Schatz, G.~C., Shelton, W.~A., Silverstein, D.~W.,
  Smith, D. M.~A., Soares, T.~A., Song, D., Swart, M., Taylor, H.~L., Thomas,
  G.~S., Tipparaju, V., Truhlar, D.~G., Tsemekhman, K., Van~Voorhis, T.,
  {V{\'a}zquez-Mayagoitia}, {\'A}., Verma, P., Villa, O., Vishnu, A.,
  Vogiatzis, K.~D., Wang, D., Weare, J.~H., Williamson, M.~J., Windus, T.~L.,
  Woli{\'n}ski, K., Wong, A.~T., Wu, Q., Yang, C., Yu, Q., Zacharias, M.,
  Zhang, Z., Zhao, Y., and Harrison, R.~J.}
\newblock {{NWChem}}: Past, present, and future.
\newblock {\em Journal of Chemical Physics 152}, 18 (May 2020), 184102.

\bibitem{bakerLocationTransitionStates1996}
{\sc Baker, J., and Chan, F.}
\newblock The location of transition states: {{A}} comparison of {{Cartesian}},
  {{Z-matrix}}, and natural internal coordinates.
\newblock {\em Journal of Computational Chemistry 17}, 7 (1996), 888--904.

\bibitem{berahasLimitedmemoryBFGSDisplacement2022}
{\sc Berahas, A.~S., Curtis, F.~E., and Zhou, B.}
\newblock Limited-memory {{BFGS}} with displacement aggregation.
\newblock {\em Mathematical Programming 194}, 1 (July 2022), 121--157.

\bibitem{burknerBrmsPackageBayesian2017}
{\sc B{\"u}rkner, P.-C.}
\newblock Brms: {{An R Package}} for {{Bayesian Multilevel Models Using Stan}}.
\newblock {\em Journal of Statistical Software 80}, 1 (Aug. 2017), 1--28.

\bibitem{cancesImprovementsActivationrelaxationTechnique2009}
{\sc Canc{\`e}s, E., Legoll, F., Marinica, M.-C., Minoukadeh, K., and Willaime,
  F.}
\newblock Some improvements of the activation-relaxation technique method for
  finding transition pathways on potential energy surfaces.
\newblock {\em Journal of Chemical Physics 130}, 11 (Mar. 2009), 114711.

\bibitem{carpenterStanProbabilisticProgramming2017}
{\sc Carpenter, B., Gelman, A., Hoffman, M.~D., Lee, D., Goodrich, B.,
  Betancourt, M., Brubaker, M., Guo, J., Li, P., and Riddell, A.}
\newblock Stan: {{A Probabilistic Programming Language}}.
\newblock {\em Journal of Statistical Software 76}, 1 (2017).

\bibitem{cerjanFindingTransitionStates1981}
{\sc Cerjan, C.~J., and Miller, W.~H.}
\newblock On finding transition states.
\newblock {\em The Journal of Chemical Physics 75}, 6 (Sept. 1981), 2800--2806.

\bibitem{chillBenchmarksCharacterizationMinima2014}
{\sc Chill, S.~T., Stevenson, J., Ruehle, V., Shang, C., Xiao, P., Farrell,
  J.~D., Wales, D.~J., and Henkelman, G.}
\newblock Benchmarks for {{Characterization}} of {{Minima}}, {{Transition
  States}}, and {{Pathways}} in {{Atomic}}, {{Molecular}}, and {{Condensed
  Matter Systems}}.
\newblock {\em Journal of Chemical Theory and Computation 10}, 12 (Dec. 2014),
  5476--5482.

\bibitem{chillEONSoftwareLong2014}
{\sc Chill, S.~T., Welborn, M., Terrell, R., Zhang, L., Berthet, J.-C.,
  Pedersen, A., J{\'o}nsson, H., and Henkelman, G.}
\newblock {{EON}}: Software for long time simulations of atomic scale systems.
\newblock {\em Modelling and Simulation in Materials Science and Engineering
  22}, 5 (July 2014), 055002.

\bibitem{eyringActivatedComplexChemical1935}
{\sc Eyring, H.}
\newblock The activated complex in chemical reactions.
\newblock {\em Journal of Chemical Physics 3}, 2 (Feb. 1935), 107--115.

\bibitem{flemingHowNotLie1986}
{\sc Fleming, P.~J., and Wallace, J.~J.}
\newblock How not to lie with statistics: The correct way to summarize
  benchmark results.
\newblock {\em Communications of the ACM 29}, 3 (Mar. 1986), 218--221.

\bibitem{fuForcesAreNot2022}
{\sc Fu, X., Wu, Z., Wang, W., Xie, T., Keten, S., {Gomez-Bombarelli}, R., and
  Jaakkola, T.}
\newblock Forces are not {{Enough}}: {{Benchmark}} and {{Critical Evaluation}}
  for {{Machine Learning Force Fields}} with {{Molecular Simulations}}, Oct.
  2022.

\bibitem{goswamiWailordParsersReproducibility2022}
{\sc Goswami, R.}
\newblock Wailord: {{Parsers}} and {{Reproducibility}} for {{Quantum
  Chemistry}}.
\newblock {\em Proceedings of the 21st Python in Science Conference\/} (2022),
  193--197.

\bibitem{goswamiDatasetBayesianHierarchical2025}
{\sc Goswami, R.}
\newblock Dataset for {{Bayesian}} hierarchical models for quantitative
  estimates for performance metrics applied to saddle search algorithms, May
  2025.

\bibitem{goswamiEfficientImplementationGaussian2025}
{\sc Goswami, R., Masterov, M., Kamath, S., {Pe{\~n}a-Torres}, A., and
  J{\'o}nsson, H.}
\newblock Efficient implementation of gaussian process regression accelerated
  saddle point searches with application to molecular reactions, May 2025.

\bibitem{goswamiEfficientImplementationGaussian2025a}
{\sc Goswami, R., Masterov, M., Kamath, S., {Pena-Torres}, A., and J{\'o}nsson,
  H.}
\newblock Efficient {{Implementation}} of {{Gaussian Process Regression
  Accelerated Saddle Point Searches}} with {{Application}} to {{Molecular
  Reactions}}.
\newblock {\em Journal of Chemical Theory and Computation\/} (July 2025).

\bibitem{goswamiHighThroughputReproducible2022}
{\sc Goswami, R., and S., R.}
\newblock High {{Throughput Reproducible Literate Phylogenetic Analysis}}.
\newblock In {\em 2022 {{Seventh International Conference}} on {{Parallel}},
  {{Distributed}} and {{Grid Computing}} ({{PDGC}})\/} (Nov. 2022),
  pp.~337--340.

\bibitem{goswamiReproducibleHighPerformance2022}
{\sc Goswami, R., S., R., Goswami, A., Goswami, S., and Goswami, D.}
\newblock Reproducible {{High Performance Computing}} without {{Redundancy}}
  with {{Nix}}.
\newblock In {\em 2022 {{Seventh International Conference}} on {{Parallel}},
  {{Distributed}} and {{Grid Computing}} ({{PDGC}})\/} (Nov. 2022),
  pp.~238--242.

\bibitem{gundeExploringPotentialEnergy2024}
{\sc Gunde, M., Jay, A., Pober{\v z}nik, M., Salles, N., Richard, N., Landa,
  G., Mousseau, N., {Martin-Samos}, L., and Hemeryck, A.}
\newblock Exploring potential energy surfaces to reach saddle points above
  convex regions.
\newblock {\em The Journal of Chemical Physics 160}, 23 (June 2024), 232501.

\bibitem{henkelmanDimerMethodFinding1999}
{\sc Henkelman, G., and J{\'o}nsson, H.}
\newblock A dimer method for finding saddle points on high dimensional
  potential surfaces using only first derivatives.
\newblock {\em The Journal of Chemical Physics 111}, 15 (Oct. 1999),
  7010--7022.

\bibitem{hermesSellaOpenSourceAutomationFriendly2022}
{\sc Hermes, E.~D., Sargsyan, K., Najm, H.~N., and Z{\'a}dor, J.}
\newblock Sella, an {{Open-Source Automation-Friendly Molecular Saddle Point
  Optimizer}}.
\newblock {\em Journal of Chemical Theory and Computation 18}, 11 (Nov. 2022),
  6974--6988.

\bibitem{heydenEfficientMethodsFinding2005}
{\sc Heyden, A., Bell, A.~T., and Keil, F.~J.}
\newblock Efficient methods for finding transition states in chemical
  reactions: {{Comparison}} of improved dimer method and partitioned rational
  function optimization method.
\newblock {\em The Journal of Chemical Physics 123}, 22 (Dec. 2005), 224101.

\bibitem{homanNoUturnSamplerAdaptively2014}
{\sc Homan, M.~D., and Gelman, A.}
\newblock The {{No-U-turn}} sampler: Adaptively setting path lengths in
  {{Hamiltonian Monte Carlo}}.
\newblock {\em The Journal of Machine Learning Research 15}, 1 (Jan. 2014),
  1593--1623.

\bibitem{huberAutomatedReproducibleWorkflows2022}
{\sc Huber, S.~P.}
\newblock Automated reproducible workflows and data provenance with {{AiiDA}}.
\newblock {\em Nature Reviews Physics 4}, 7 (July 2022), 431--431.

\bibitem{jensenIntroductionComputationalChemistry2017}
{\sc Jensen, F.}
\newblock {\em Introduction to Computational Chemistry}, third edition~ed.
\newblock Wiley, Chichester, UK ; Hoboken, NJ, 2017.

\bibitem{kastnerSuperlinearlyConvergingDimer2008}
{\sc K{\"a}stner, J., and Sherwood, P.}
\newblock Superlinearly converging dimer method for transition state search.
\newblock {\em The Journal of Chemical Physics 128}, 1 (Jan. 2008), 014106.

\bibitem{koistinenMinimumModeSaddle2020}
{\sc Koistinen, O.-P., {\'A}sgeirsson, V., Vehtari, A., and J{\'o}nsson, H.}
\newblock Minimum {{Mode Saddle Point Searches Using Gaussian Process
  Regression}} with {{Inverse-Distance Covariance Function}}.
\newblock {\em Journal of Chemical Theory and Computation 16}, 1 (Jan. 2020),
  499--509.

\bibitem{kriegelBlackArtRuntime2017}
{\sc Kriegel, H.-P., Schubert, E., and Zimek, A.}
\newblock The (black) art of runtime evaluation: {{Are}} we comparing
  algorithms or implementations?
\newblock {\em Knowledge and Information Systems 52}, 2 (Aug. 2017), 341--378.

\bibitem{kuffnerWhyArePValues2019}
{\sc Kuffner, T.~A., and Walker, S.~G.}
\newblock Why are p-{{Values Controversial}}?
\newblock {\em The American Statistician 73}, 1 (Jan. 2019), 1--3.

\bibitem{lengEfficientSoftestMode2013}
{\sc Leng, J., Gao, W., Shang, C., and Liu, Z.-P.}
\newblock Efficient softest mode finding in transition states calculations.
\newblock {\em Journal of Chemical Physics 138}, 9 (Mar. 2013), 94110.

\bibitem{mcelreathStatisticalRethinkingBayesian2020}
{\sc McElreath, R.}
\newblock {\em Statistical Rethinking: A {{Bayesian}} Course with Examples in
  {{R}} and {{Stan}}}, 2~ed.
\newblock {{CRC}} Texts in Statistical Science. {Taylor and Francis, CRC
  Press}, Boca Raton, 2020.

\bibitem{mcshaneAbandonStatisticalSignificance2019}
{\sc McShane, B.~B., Gal, D., Gelman, A., Robert, C., and Tackett, J.~L.}
\newblock Abandon {{Statistical Significance}}.
\newblock {\em The American Statistician 73}, sup1 (Mar. 2019), 235--245.

\bibitem{melanderRemovingExternalDegrees2015}
{\sc Melander, M., Laasonen, K., and J{\'o}nsson, H.}
\newblock Removing {{External Degrees}} of {{Freedom}} from {{Transition-State
  Search Methods}} using {{Quaternions}}.
\newblock {\em Journal of Chemical Theory and Computation 11}, 3 (Mar. 2015),
  1055--1062.

\bibitem{molderSustainableDataAnalysis2021}
{\sc M{\"o}lder, F., Jablonski, K.~P., Letcher, B., Hall, M.~B.,
  {Tomkins-Tinch}, C.~H., Sochat, V., Forster, J., Lee, S., Twardziok, S.~O.,
  Kanitz, A., Wilm, A., Holtgrewe, M., Rahmann, S., Nahnsen, S., and
  K{\"o}ster, J.}
\newblock Sustainable data analysis with {{Snakemake}}, Apr. 2021.

\bibitem{mousseauTravelingPotentialEnergy1998}
{\sc Mousseau, N., and Barkema, G.~T.}
\newblock Traveling through potential energy landscapes of disordered
  materials: The activation-relaxation technique.
\newblock {\em Physical Review E 57}, 2 (Feb. 1998), 2419--2424.

\bibitem{mousseauActivationRelaxationTechniqueART2012}
{\sc Mousseau, N., B{\'e}land, L.~K., Brommer, P., Joly, J.-F., {El-Mellouhi},
  F., {Machado-Charry}, E., Marinica, M.-C., and Pochet, P.}
\newblock The {{Activation-Relaxation Technique}}: {{ART Nouveau}} and
  {{Kinetic ART}}.
\newblock {\em Journal of Atomic and Molecular Physics 2012}, 1 (2012), 925278.

\bibitem{munroDefectMigrationCrystalline1999}
{\sc Munro, L.~J., and Wales, D.~J.}
\newblock Defect migration in crystalline silicon.
\newblock {\em Physical Review B 59}, 6 (Feb. 1999), 3969--3980.

\bibitem{nicholsWalkingPotentialEnergy1990}
{\sc Nichols, J., Taylor, H., Schmidt, P., and Simons, J.}
\newblock Walking on potential energy surfaces.
\newblock {\em Journal of Chemical Physics 92}, 1 (Jan. 1990), 340--346.

\bibitem{nocedalNumericalOptimization2006}
{\sc Nocedal, J., and Wright, S.~J.}
\newblock {\em Numerical Optimization}, 2nd ed~ed.
\newblock Springer Series in Operations Research. Springer, New York, 2006.

\bibitem{oharaNotLogtransformCount2010}
{\sc O'Hara, R.~B., and Kotze, D.~J.}
\newblock Do not log-transform count data.
\newblock {\em Methods in Ecology and Evolution 1}, 2 (2010), 118--122.

\bibitem{olsenComparisonMethodsFinding2004}
{\sc Olsen, R.~A., Kroes, G.~J., Henkelman, G., Arnaldsson, A., and
  J{\'o}nsson, H.}
\newblock Comparison of methods for finding saddle points without knowledge of
  the final states.
\newblock {\em The Journal of Chemical Physics 121}, 20 (Nov. 2004),
  9776--9792.

\bibitem{pechukasRecentDevelopmentsTransition1982}
{\sc Pechukas, P.}
\newblock Recent developments in transition state theory.
\newblock {\em Berichte der Bunsen-Gesellschaft f{\"u}r Physikalische Chemie
  86}, 5 (1982), 372--378.

\bibitem{poberznikPARTnPluginImplementation2024}
{\sc Poberznik, M., Gunde, M., Salles, N., Jay, A., Hemeryck, A., Richard, N.,
  Mousseau, N., and {Martin-Samos}, L.}
\newblock {{pARTn}}: {{A}} plugin implementation of the {{Activation Relaxation
  Technique}} nouveau that takes over the {{FIRE}} minimisation algorithm.
\newblock {\em Computer Physics Communications 295\/} (Feb. 2024), 108961.

\bibitem{sallermannFlowyHighPerformance2024}
{\sc Sallermann, M., Goswami, A., {Pe{\~n}a-Torres}, A., and Goswami, R.}
\newblock Flowy: High performance probabilistic lava emplacement prediction,
  June 2024.

\bibitem{starkBenchmarkingMachineLearning2024}
{\sc Stark, W.~G., {van der Oord}, C., Batatia, I., Zhang, Y., Jiang, B.,
  Cs{\'a}nyi, G., and Maurer, R.~J.}
\newblock Benchmarking of machine learning interatomic potentials for reactive
  hydrogen dynamics at metal surfaces, Mar. 2024.

\bibitem{talirzMaterialsCloudPlatform2020}
{\sc Talirz, L., Kumbhar, S., Passaro, E., Yakutovich, A.~V., Granata, V.,
  Gargiulo, F., Borelli, M., Uhrin, M., Huber, S.~P., Zoupanos, S., Adorf,
  C.~S., Andersen, C.~W., Sch{\"u}tt, O., Pignedoli, C.~A., Passerone, D.,
  VandeVondele, J., Schulthess, T.~C., Smit, B., Pizzi, G., and Marzari, N.}
\newblock Materials cloud, a platform for open computational science.
\newblock {\em Scientific Data 7}, 1 (Sept. 2020), 299.

\bibitem{walesEnergyLandscapesClusters2000}
{\sc Wales, D.~J., Doye, J. P.~K., Miller, M.~A., Mortenson, P.~N., and Walsh,
  T.~R.}
\newblock Energy landscapes: From clusters to biomolecules.
\newblock In {\em Advances in {{Chemical Physics}}}. John Wiley \& Sons, Ltd,
  2000, pp.~1--111.

\bibitem{zengUnificationAlgorithmsMinimum2014}
{\sc Zeng, Y., Xiao, P., and Henkelman, G.}
\newblock Unification of algorithms for minimum mode optimization.
\newblock {\em The Journal of Chemical Physics 140}, 4 (Jan. 2014), 044115.

\end{thebibliography}
